\documentclass[final,3p,times]{elsarticle}

\usepackage{graphicx}

\usepackage{amssymb}
\usepackage{amsthm}

\usepackage{amsmath}
\usepackage{amsfonts}
\usepackage{hyperref}
\usepackage{float}
\usepackage{color,soul}
\usepackage{xcolor}
\usepackage[normalem]{ulem}

\allowdisplaybreaks

\newcounter{bla}

\journal{Computer Physics Communications}

\begin{document}

\begin{frontmatter}



\title{Simulating ultrarelativistic beam-plasma instabilities with a quasistatic particle-in-cell code}


\author[a,b]{Q. Labro\corref{author}}
\author[a,b]{X. Davoine}
\author[a,b]{L. Gremillet}
\author[a,b,c]{L. Bergé}

\cortext[author] {Corresponding author.\\\textit{E-mail address:} quentin.labro@cea.fr}
\address[a]{CEA, DAM, DIF, F-91297 Arpajon, France}
\address[b]{Université Paris-Saclay, CEA, LMCE, F-91680 Bruyères-le-Châtel, France}
\address[c]{Centre Lasers Intenses et Applications, Université de Bordeaux-CNRS-CEA, UMR 5107, 33405 Talence, France}

\begin{abstract}

Quasistatic particle-in-cell (PIC) codes are increasingly employed to study laser or plasma wakefield accelerators. By decoupling the slow dynamics of the driver (a laser or ultrarelativistic particle beam) from the fast plasma response, these codes can reduce the computational time by several orders of magnitude compared to conventional PIC codes. In this work, we demonstrate that quasistatic PIC codes can also be utilized to investigate relativistic beam-plasma instabilities, with a focus on the oblique two-stream instability (OTSI). For this purpose, we have developed a 2D quasistatic PIC code, QuaSSis, based on a new numerical scheme that can handle transversely periodic boundary conditions, a capability absent in previous quasistatic codes.
The accuracy of QuaSSis is benchmarked first against standard PIC simulations performed with
the CALDER code, and then against an analytical spatiotemporal model of the OTSI. Physically, this instability grows exponentially from initial fluctuations in the particle charge or current densities. Since the numerical noise inherent to PIC simulations can mimic these fluctuations to some extent, its control is crucial to seed the beam-plasma instability at the desired amplitude. Common methods for tuning this noise involve modifying the resolution or adding filters, but these can be computationally costly when aiming at
very low noise levels. Here, we show that this noise can be finely controlled by properly initializing the positions and weights of the macroparticles.

\end{abstract}

\begin{keyword}
particle-in-cell simulations; quasistatic method; beam-plasma instabilities.
\end{keyword}

\end{frontmatter}


\section{Introduction}
\label{Introduction}

Understanding the interaction of relativistic particle beams from active galactic nuclei, gamma-ray bursts, and pulsars with their surrounding intergalactic or interstellar environments is a key focus of high-energy astrophysics \cite{Kumar_2015}. This interaction is often mediated by collective processes, and in particular by micro-instabilities arising from electromagnetic fluctuations or thermal noise at kinetic scales \cite{Davidson_1983, BretInstaRegime}. These instabilities can amplify the initial electromagnetic fields by several orders of magnitude, until they profoundly alter the phase space of the beam, leading to their saturation \cite{Davidson_1972, Thode_1975, Kato_2005}. The generated electromagnetic fields can deflect the beam particles, resulting in the formation of collisionless shocks \cite{Marcowith_2016, Lemoine_PRL_2019} or the emission of broad spectra of high-energy photons via synchrotron radiation \cite{Piran_2005, Meszaros_2006}.

These phenomena can also be explored in laboratory experiments by directing an electron beam, produced either by a conventional accelerator \cite{Allen_2012} or an intense laser pulse \cite{Tatarakis_2003}, into a plasma target. Recently, it has even been proposed to harness them to produce bright gamma-ray sources \cite{Benedetti_2018}.

This paper addresses the modeling of the unstable propagation of ultrarelativistic electron or electron-positron pair beams through much denser, initially unmagnetized plasmas. Micro-instabilities in these systems can be categorized by the orientation of their wave vectors and the nature (electrostatic or electromagnetic) of the driven fluctuations. Specifically, the two-stream instability (TSI) generates purely electrostatic fluctuations in the longitudinal direction \cite{Bohm&Gross, Fainberg_1970, Rudakov_1971}, while the current filament instability (CFI), a variant of the Weibel instability fed by momentum anisotropies \cite{Weibel_1959}, amplifies essentially magnetic modulations in the transverse direction \cite{Fried_1959, Califano_1997}. In addition, the oblique two-stream instability (OTSI), sometimes viewed as a mix of the TSI and CFI \cite{BretInstaRegime}, generates mainly electrostatic modulations propagating at oblique angles.

Particle-in-cell (PIC) codes \cite{Birdsall_2004} are primary tools for simulating these kinetic processes. Yet they struggle with ultrarelativistic beams because the timescale of the beam evolution, determined by the instability growth time, is much longer than the plasma period of the background plasma electrons. The problem is that this plasma period, and the associated plasma skin depth, must be well resolved to accurately describe the interaction (and ensure the stability of the simulation). This constraint can entail very expensive computations even though the beam and plasma profiles remain nearly static in the beam's comoving frame at each time step.
For example, modeling beam-plasma instabilities in millimeter-thick solid targets, as envisioned at SLAC's FACET-II 10~GeV accelerator facility \cite{Yakimenko_2019, Corde_2019, PSMC_2022}, requires hundreds of thousands of attosecond-scale time steps. Consequently, standard PIC simulations are often limited to simplified 2D geometries, unless extremely expensive 3D calculations are undertaken.

In this paper, we show that ultrarelativistic beam-plasma instabilities can be efficiently modeled using the quasistatic PIC (QS-PIC) method \cite{Mora_1997}. Successfully employed for decades in the design of plasma wakefield accelerators (PWFAs) \cite{Chen_1985} and laser wakefield accelerators (LWFAs) \cite{Faure_2004}, this method assumes that the driving beam negligibly changes during the transit time of a plasma particle. In this two-timescale approach, implemented in the beam's comoving frame, the trajectories of the fast-evolving plasma electrons can be computed with good accuracy while treating the beam as stationary. This simplification allows the plasma response to be described only over the long timescale of beam evolution, dramatically reducing the number of time steps and computational cost.

Several QS-PIC codes have already been developed in past decades, including WAKE \cite{Mora_1997}, QuickPIC \cite{QuickPIC_2006, QuickPIC_2013}, Lcode \cite{LCode_1999, LCODE_2016}, HiPACE \cite{HiPACE_2014}, HiPACE++ \cite{HiPACE_2022}, WAND-PIC \cite{WAND_2017, WANPIC_2022}, and QPAD \cite{QPAD_2021}. They have demonstrated simulation time reductions of several orders of magnitude compared to conventional PIC codes. However, to our knowledge, their use has so far been limited to PWFA and LWFA scenarios.

The disparity of the beam and plasma time scales leveraged in QS-PIC simulations of PWFAs also holds for the instability-governed interaction of relativistic beams with dense plasmas,
making the QS-PIC technique very attractive for tackling this problem. Yet several issues must be addressed. The first concerns the benchmark of the QS-PIC results against instability theories, which typically assume transversely uniform systems. These can only be simulated by periodic transverse boundary conditions but, to our knowledge, such conditions have never been implemented in QS-PIC codes. 
Second, precise control of the initial fluctuations seeding the instabilities is needed.
Third, the evolution of the beam must be tracked over large enough propagation distances to characterize the growth rate of the dominant instability and observe the onset of the saturation stage.
The purpose of this work is to show that relativistic beam-plasma instabilities, and specifically the OTSI, can be accurately simulated over long interaction distances by a QS-PIC code with the ability to handle periodic transverse boundary conditions and to control the initial noise level.

The paper is organized as follows. First, we review the fundamentals of the OTSI and illustrate them through a standard PIC simulation using the CALDER code \cite{Lefebvre_2003}. Next, we introduce the general framework of our new quasistatic PIC code, QuaSSis, and validate it against CALDER for OTSI simulations. We then detail the implementation of periodic transverse boundary conditions and compare the simulation results with an analytical model describing the spatiotemporal growth of the OTSI \cite{PSMC_2022}. Finally, we propose various methods to control numerical noise in QS-PIC simulations and discuss their performance.

\section{Beam-plasma instabilities and quasistatic approximation}
\label{Sect1}

\subsection{Theoretical description}

Beam-plasma instabilities typically evolve through two distinct phases. The initial linear phase is characterized by the exponential amplification of coupled modulations in the field and particle distributions. This phase lasts until the effect of the fields on the particles is no longer perturbative, at which point the field growth stops, or greatly slows down. The instability then enters its nonlinear phase, generally marked by slower-evolving, large-amplitude modulations in the field and particle distributions.

In this study, we focus solely on the linear phase of the unstable interaction of an ultrarelativistic beam (with electron density $n_b$ and Lorentz factor $\gamma_b \gg 1$) with a much denser plasma (with electron density $n_p \gg n_b$).
We recall in Table~\ref{table:growth_rate} the maximum temporal growth rates for the CFI \cite{Godfrey_1975}, TSI \cite{Bludman_1960} and OTSI \cite{Watson_1960} in the cold-fluid limit and for uniform infinite systems.
We have introduced the background plasma frequency $\omega_p = \sqrt{n_p e^2/m_e \varepsilon_0}$ and the corresponding plasma wave number $k_p \equiv c/\omega_p$, with $e$ as the elementary charge, $m_e$ the electron mass, $c$ the velocity of light, and $\epsilon_0$ the vacuum permittivity. The CFI and OTSI growth rates also involve the transverse wave number $k_\perp$.

\begin{table*}
\centering
\begin{tabular}{|l|c|c|c|}
  \hline
    Instability & CFI & TSI & OTSI  \\
  \hline
    Growth rate & $\left( \frac{1}{\gamma_b} \frac{n_b}{n_p} \frac{k_\perp^2}{k_p^2 + k_\perp^2} \right)^{1/2} \omega_p$ 
    & $\frac{\sqrt{3}}{2^{4/3}} \frac{1}{\gamma_b} \left( \frac{n_b}{n_p} \right)^{1/3} \omega_p$ 
    & $\frac{\sqrt{3}}{2^{4/3}} \left( \frac{1}{\gamma_b} \frac{n_b}{n_p} \frac{k_\perp^2}{k_p^2 + k_\perp^2}\right)^{1/3} \omega_p$ \\
  \hline

\end{tabular}
    \caption{Temporal growth rates of the CFI \cite{Godfrey_1975}, TSI \cite{Bludman_1960}, and OTSI \cite{Watson_1960} for infinite, uniform, cold-fluid beam-plasma systems in the ultrarelativistic ($\gamma_b \gg 1$) and dilute ($n_b \ll n_p$) beam limit. 
    }
    \label{table:growth_rate}
\end{table*}

From Table~\ref{table:growth_rate}, a hierarchy can be established between the instabilities \cite{BretHierarchyInsta}, given that the TSI growth rate scales as $\gamma_b^{-1}$, the CFI growth rate as $\gamma_b^{-1/2}$, and the OTSI growth rate as $\gamma_b^{-1/3}$. Consequently, the OTSI dominates for beam Lorentz factors $\gamma_b \gg 1$ and density ratios $\alpha \equiv n_b/n_p \ll 1$. All simulations presented in this study are run within this OTSI-dominated regime.

Early beam-plasma instability theories assumed infinite, uniform beam-plasma systems, resulting in purely temporal (i.e. space-independent) growth of the particle and field perturbations at the rates presented in Table~\ref{table:growth_rate}. However, in the more realistic case of a beam with a steep longitudinal boundary, spatiotemporal effects can arise near the beam front, as explored a long time ago for the TSI \cite{Briggs_1964, Bers_1983, Jones_1983}.  
Spatiotemporal models have since been developed for the relativistic CFI \cite{Pathak_2015} and OTSI \cite{PSMC_2022}. Such models are essential for describing laboratory experiments involving ultrashort beams.

For a relativistic particle beam propagating along $x > 0$ at a velocity $v_x \simeq c$, the derivation of the spatiotemporal OTSI model \cite{PSMC_2022} is simplified by changing to comoving coordinates:
\begin{equation}
    \xi = c t - x \quad \text{ and } \quad s = x \,.
\end{equation}
Here, $\xi$ represents the longitudinal position relative to the beam front and $s$ is the propagation distance.
The lab-frame derivatives then transform as 
\begin{equation}
    \frac{\partial}{\partial t} = c\frac{\partial}{\partial \xi}
    \quad \text{ and } \quad
    \frac{\partial}{\partial x} = \frac{\partial}{\partial s} - \frac{\partial}{\partial \xi} \,.
    \label{eq:qs_derivatives}
\end{equation}
The quasistatic approximation assumes that the plasma and the field profiles evolve much more rapidly with $\xi$ than the beam varies with $s$, meaning $\partial_\xi \gg \partial_s$. Thus,
\begin{equation}
    \frac{\partial}{\partial t} = c\frac{\partial}{\partial \xi} 
    \quad \text{ and } \quad
    \frac{\partial}{\partial x} \simeq - \frac{\partial}{\partial \xi} \,.
\end{equation}
Combining the linearized cold-fluid equations for the beam and plasma electrons with Poisson's equation, and  applying the slowly varying envelope approximation, one can obtain the equation governing the envelope ($\delta E_y$) of the transverse electric field 
\begin{equation}
    \left[\partial_s^2 \partial_\xi + \mathrm{i} \left(\frac{2}{\sqrt{3}}\frac{\Gamma_{\rm OTSI}}{c}\right)^3 \right] \delta E_y = 0 \,,
\end{equation}
where $\Gamma_{\rm OTSI}$ is the (temporal) OTSI growth rate given in Table~\ref{table:growth_rate} and $\rm i$ is the imaginary unit. The total perturbed field is given by $E_y^{(1)} =  \delta E_y e^{-\mathrm{i} k_p \xi + \mathrm{i} \mathbf{k_\perp} \cdot \mathbf{r_\perp}}$. 
The impulsive solution (i.e. with a Dirac function source at $\xi=s=0$) to this equation behaves asymptotically in $s$ as \cite{PSMC_2022}
\begin{equation}
    \delta E_y (\xi, s, k_\perp) \propto
    \exp{\left( 
    \frac{\sqrt{3}}{2^{2/3}c} \left(\sqrt{3}+\mathrm{i} \right) \Gamma_{\rm OTSI} \xi^{1/3} s^{2/3}
    - \mathrm{i} \frac{\pi}{12}
    \right)} \,.
    \label{eq:Gamma_spt}
\end{equation}
This formula describes the spatiotemporal growth of the OTSI as a function of $\xi$ and $s$ for $s \gg \xi$.

\subsection{Simulation of the OTSI with a standard PIC code}
 
The main motivation of this paper is to demonstrate that ultrarelativistic beam-plasma instabilities can be accurately simulated using the QS-PIC technique. For this purpose, we first perform a reference simulation with the standard PIC code CALDER \cite{Lefebvre_2003} and detail our analysis of the OTSI dynamics. This simulation uses a neutral electron-positron pair beam to avoid the excitation of a plasma wakefield or the generation of a return current (as would be the case with a nonneutral beam), which can complicate the analysis of the interaction.
Specifically, we consider a $\sim$10~GeV ($\gamma_b = 20\,000$) monoenergetic beam with a density profile that is uniform in the transverse direction and flat-top in the longitudinal direction, with a total (electrons and positrons) density of  $n_b = 3 \times 10^{18}\,\rm cm^{-3}$ and a length of $75.21\,\rm \mu m$ ($100\,k_p^{-1}$). The background plasma density is $n_p = 5 \times 10^{19}\,\rm cm^{-3}$, corresponding to a density ratio $n_b/n_p \equiv \alpha = 0.06$, and the plasma temperature is $T_p=0.01\,\rm eV$. 
The simulation is performed in a 2D3V geometry (two-dimensional in space, three-dimensional in momentum space). The spatial domain has dimensions $L_x\times L_y = 91.2 \times 37.6\,\rm \mu m^2$ ($120 \times 50\,k_p^{-2}$). The mesh size is $\Delta x = \Delta y = 0.038\,\rm \mu m$ ($0.05\,k_p^{-1}$) and the time step is $\Delta t = 0.088\,\rm fs$ ($0.035\,\rm\omega_p^{-1}$). 
The beam is centered in the domain, and a moving window is used to follow its propagation through the plasma. Its transverse profile extends over the full width of the box, and periodic boundary conditions are applied to both fields and particles in the transverse direction, mimicking a transversely infinite beam.  Each population (plasma electrons, beam electrons and beam positrons) is modeled by 10 macroparticles (MPs) per cell. As will be explained in Sec.~\ref{sect_noise}, the numerical weight of each MP is randomly modified to control the initial noise. To prevent a violent early evolution of the beam due to (i) the neglect of its self-fields in the vacuum and (ii) the perturbation induced by its entry into the plasma, the beam is first made to propagate ballistically until the plasma response becomes quasistatic. In the following, the time origin ($s=0$) is when the beam motion ceases to be ballistic.

\begin{figure}
    \centering
    \includegraphics[scale=0.35]{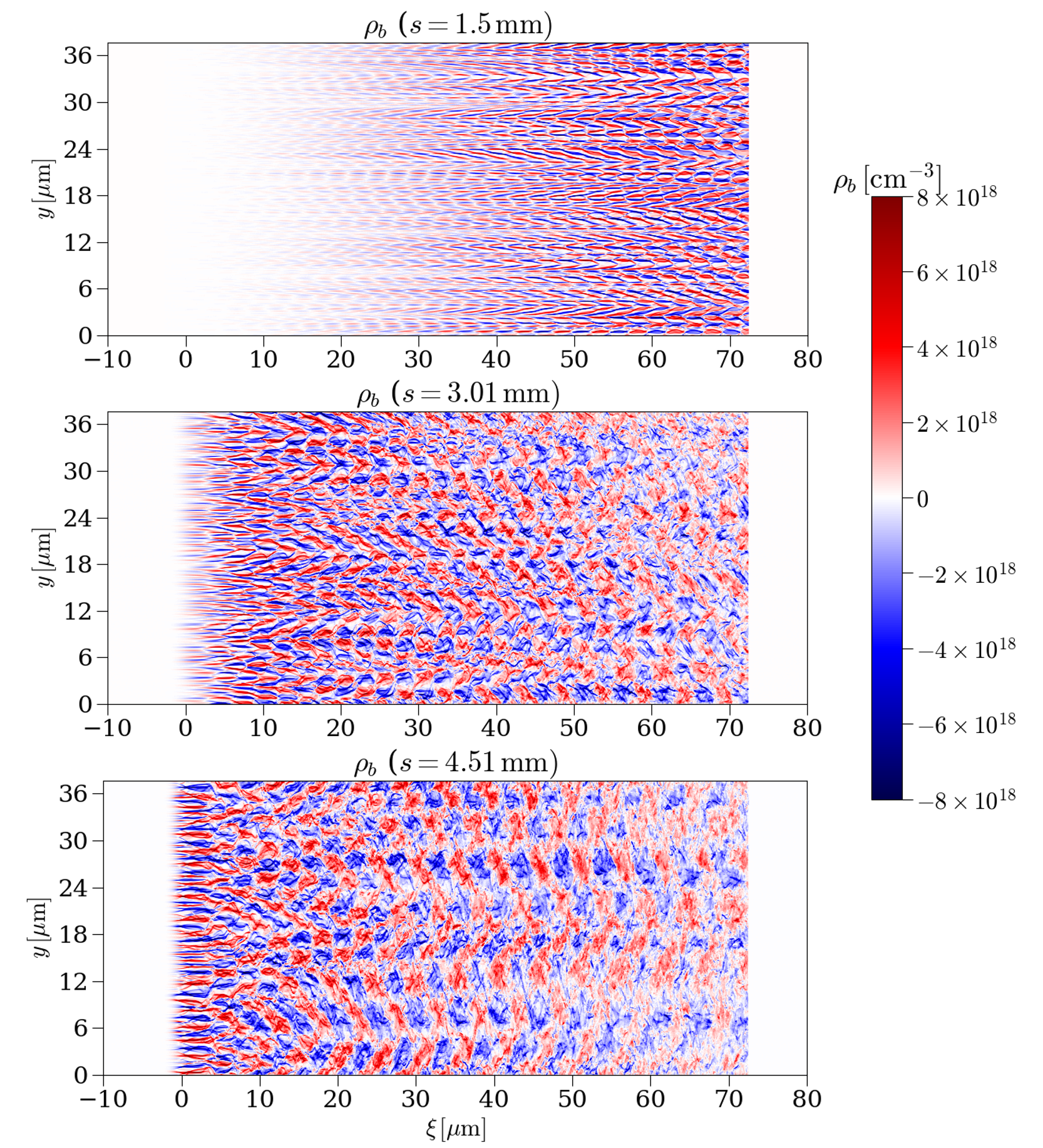} 
    \caption{Standard PIC simulation of the OTSI: charge density map of a $10\,\rm GeV$ neutral pair (electron-positron) beam propagating in a plasma with a density of $n_p = 5 \times 10^{19}\,\rm cm^{-3}$ at three different propagation distances $s\in (1.5,3.01,4.51)\,\rm mm$.
    The total beam density (electrons and positrons) is $n_b = 3 \times 10^{18}\,\rm cm^{-3}$ ($\alpha=0.06$). The initial flat-top profile of the beam extends from $\xi =0\,\rm \mu m$ (beam front) to $\xi=75.21\,\rm \mu m$. Periodic boundary conditions are used in the transverse ($y$) direction.}
\label{fig:CLP_Calder}
\end{figure}

Figure~\ref{fig:CLP_Calder} shows the beam charge density after different propagation distances. At $s=1.5\,\rm mm$, oblique modulations, with a pattern characteristic of the OTSI \cite{PSMC_2022}, are seen to grow exponentially from the front to about the middle of the beam ($\xi \simeq 45\,\rm \mu m$). Further away, their spatial growth is slower because they have nearly reached their saturation level. At $s=3.01\,\rm mm$, the OTSI is saturated within most of the beam, except near the front ($\xi \lesssim 10\,\rm \mu m$) where it is still growing. Finally, at $s=4.51\,\rm mm$, the instability has saturated throughout the beam, while nonlinear effects have started altering the shape and spectrum of the density modulations.

\begin{figure}
    \centering
    \includegraphics[scale=0.25]{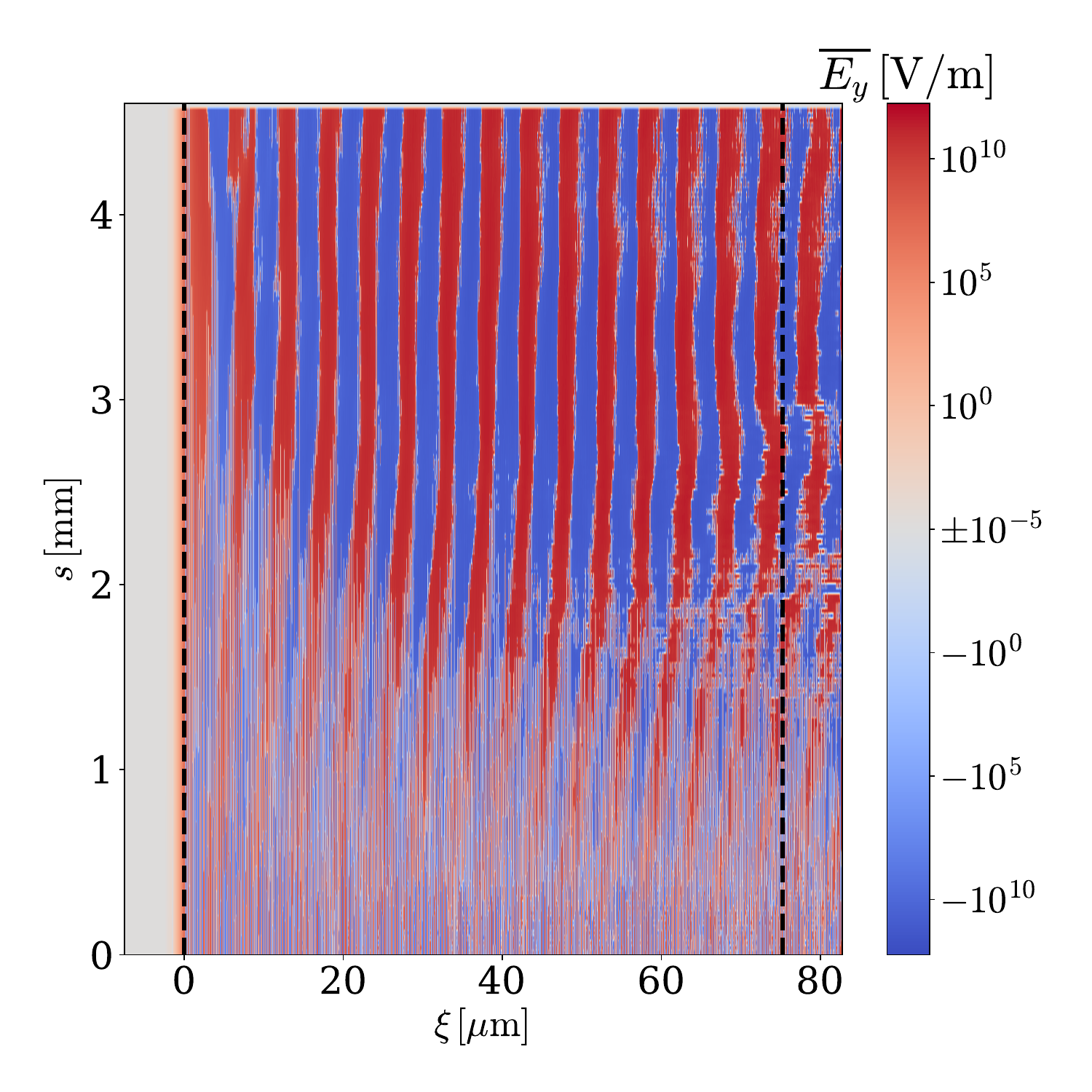}
    \caption{Evolution of the transverse $E_y$ field at the center of the beam ($y=18.8\,\rm\mu m$) as a function of $\xi$ and $s$. The field amplitude is plotted using a signed logarithmic scale:  $\overline{E_y} = {\rm sign}(E_y) \log_{10}\left[ {\rm max}(10^{-5},|E_y|)\right]$.}
    \label{fig:colomban}
\end{figure}

To confirm that the interaction verifies the quasistatic approximation, we plot in Fig.~\ref{fig:colomban} the transverse electric field ($E_y$) at the center of the beam ($y=18.8\,\rm\mu m$) as a function of $\xi$ and $s$. To accommodate the exponential growth of the field due to the instability, the figure shows $\overline{E}_y = {\rm sign}(E_y) \log_{10}\left[{\rm max}(10^{-5},|E_y|)\right]$, a transformation that displays $E_y$ in logarithmic scale while preserving the sign of the field.

At the beginning of the interaction (small $s$), only noise is observed. Then, modulations along $\xi$ with a wavelength close to the plasma wavelength $\lambda_p=2\pi/k_p$ appear, as expected from the OTSI theory. The logarithmic colorscale highlights the exponential growth of the OTSI as a function of the propagation distance $s$, up to the saturation, which is reached when the amplitude of the $E_y$ modulations stays nearly constant with $s$. The spatiotemporal behavior of the OTSI is also evident: modulations at small $\xi$ (beam front) grow more slowly and saturate later than at large $\xi$ (beam backside). An important result from Fig.~\ref{fig:colomban} is the contrast between the fast ($\lambda_p$-periodic) oscillations of $E_y$ with $\xi$ and the slow increase in its amplitude with $s$. This observation justifies the quasistatic approximation ($\partial_s \ll \partial_\xi$) for this set of parameters, and more generally for the study of the ultrarelativistic OTSI.

\begin{figure*}
    \centering
    \includegraphics[scale=0.3]{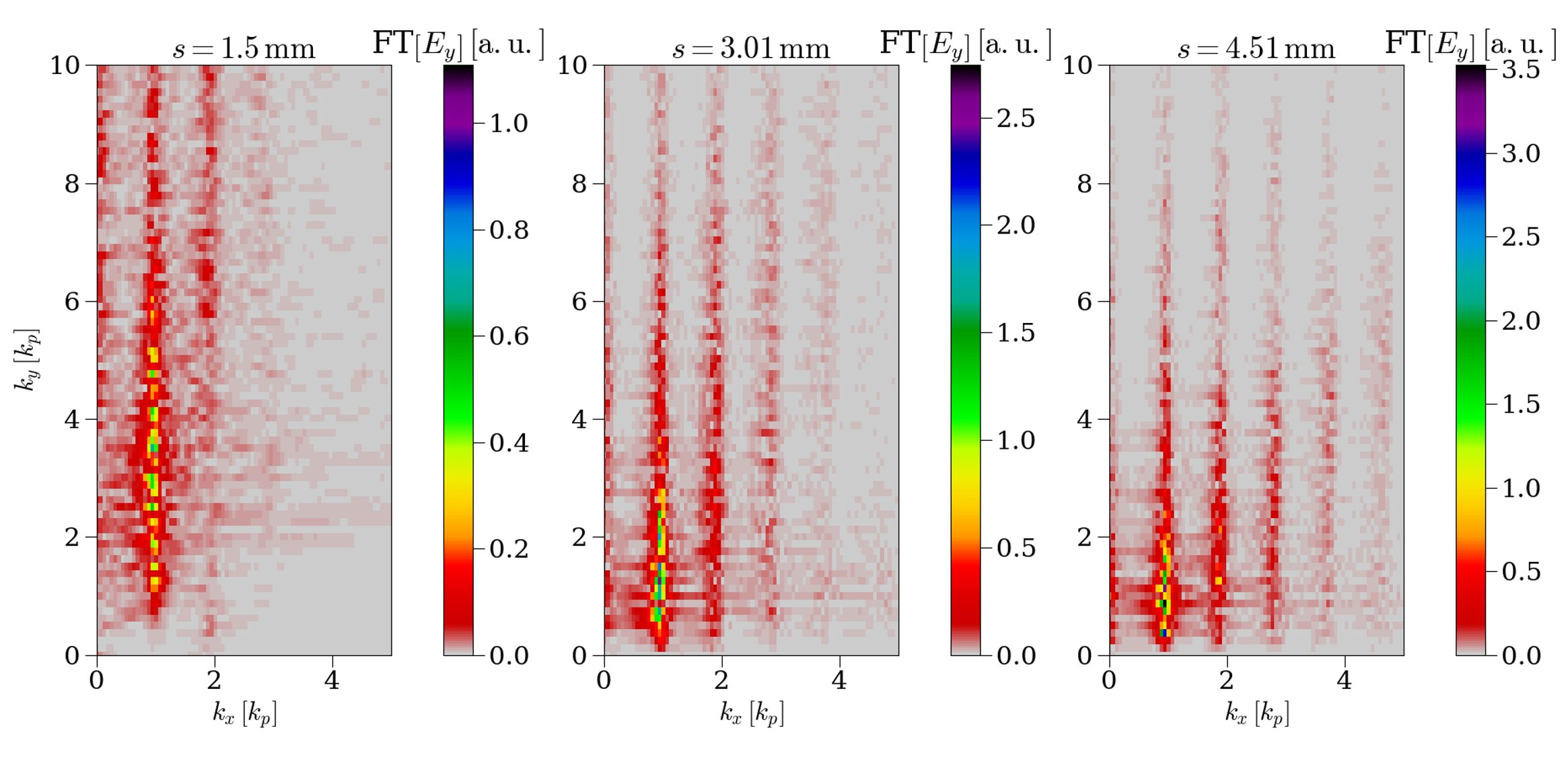}
    \caption{Norm of the Fourier transform of $E_y$ at different propagation distances. See Fig.~\ref{fig:CLP_Calder} for the corresponding charge density maps of the beam electrons.}
    \label{fig:Fourier} 
\end{figure*}

The theoretical growth rate of the dominant OTSI mode depends on its transverse wave number $k_\perp$, see Eq.~\eqref{eq:Gamma_spt}. Figure~\ref{fig:Fourier} displays the $k_x$--$k_y$ Fourier spectra of $E_y(\xi,y)$ ($k_x$ being the wavenumber associated with $\xi$) at three propagation distances. Early in the interaction ($s=1.5\,\rm mm$), the spectrum aligns with theoretical predictions \cite{BretInstaRegime}: it mainly consists of a continuum of modes centered at $k_x \simeq k_p$ and extending from $k_y \simeq k_p$ to large values, with spectral energy concentrated in the range $1 \lesssim k_y/k_p \lesssim 5$.
A harmonic replica of this continuum, indicative of the instability's nonlinearity, is also visible. At larger distances ($s=3.01\,\rm mm$ and $s=4.51\,\rm mm$), in the nonlinear stage of the instability, the $k_y$ range of the dominant modes shrinks towards $k_p$ (or even below), and higher-order harmonic replicas emerge. This behavior is consistent with the increasingly large and less oblique beam modulations observed over time in Fig.~\ref{fig:colomban}. Consequently, we will take $k_\perp = k_p$ when evaluating the theoretical growth rate in Eq.~\eqref{eq:Gamma_spt}.

\begin{figure}
    \centering
    \includegraphics[scale=0.25]{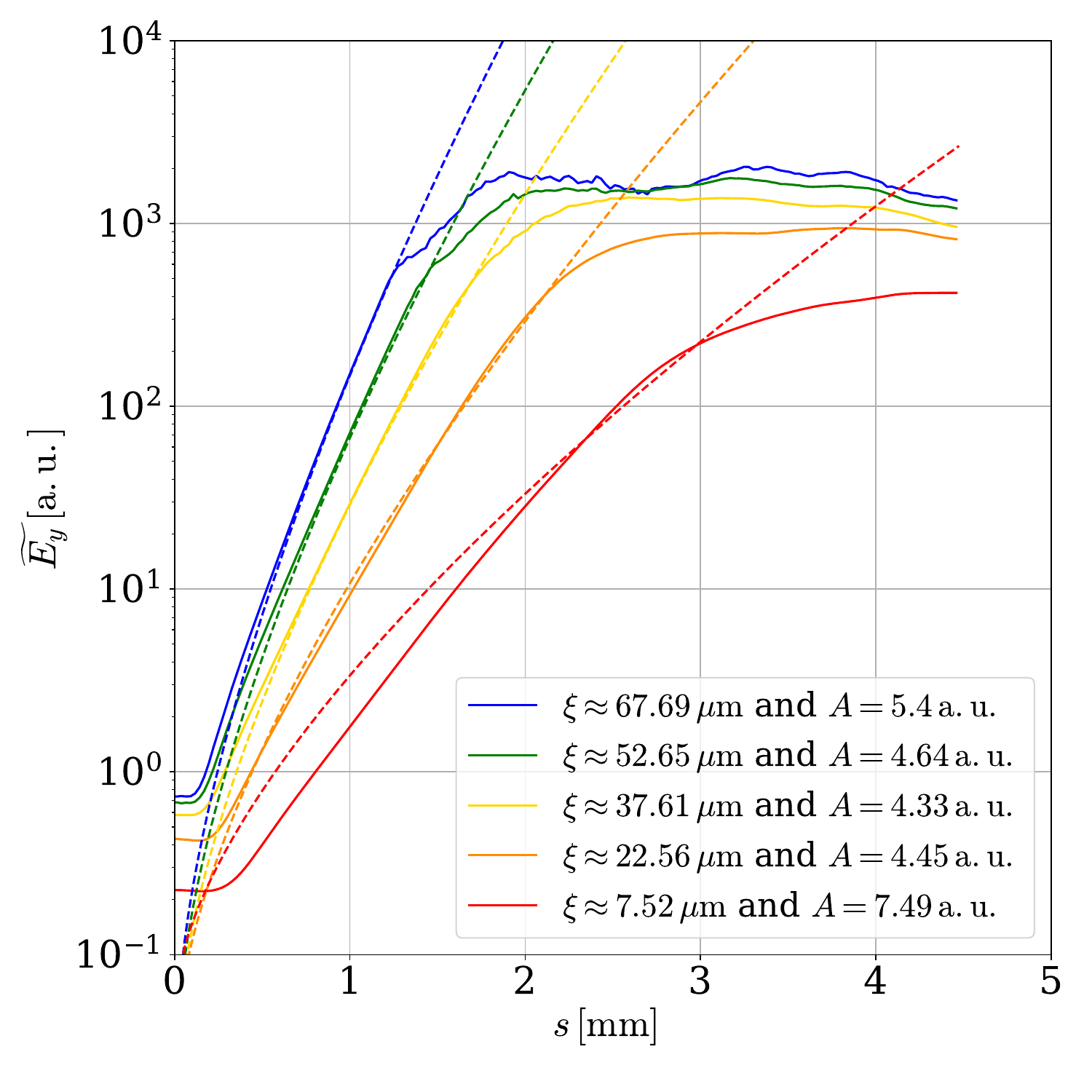}
    \caption{Evolution of the spectral amplitude $\widetilde{E}_y$ as a function of the propagation distance $s$ for various positions $\xi$ in the beam (solid curves). The dashed curves plot the theoretical prediction Eq.~\eqref{eq:Gamma_spt} with a prefactor $A$ adjusted to match the simulation data. 
    }
    \label{fig:gr_flattop}
\end{figure}

To characterize the instability dynamics as a function of $\xi$ and $s$, we perform a Fourier transform of $E_y$ over a restricted window ($10\,k_p^{-1}$ long) around various $\xi$ values. We then integrate the resulting spectrum over the spectral range $0.7 \le k_x/k_p \le 1.3$ and $0.8 \le k_y/k_p \le 3.5$ that encompasses the dominant modes. This yields the Fourier amplitude
\begin{equation}
  \widetilde{E}_y(\xi,\tau) = \frac{1}{2\pi} \int_{k_y=0.8}^{k_y=3.5}\int_{k_x=0.7}^{k_x=1.3} \left| \int_{\xi-5k_p^{-1}}^{\xi+5k_p^{-1}} \int E_y(\xi',y,\tau) \exp[\mathrm{i} (k_x\xi'+k_yy)] \,dy d\xi' \right| dk_x dk_y \,.
  \label{E_y_tilde}
\end{equation}
Figure~\ref{fig:gr_flattop} plots $\widetilde{E}_y$ as a function of $s$ for different positions $\xi$ (solid curves), revealing a clear trend of increasingly fast field growth as one moves further from the beam front. The dashed curves in Fig.~\ref{fig:gr_flattop} represent the expected spatiotemporal growth of the OTSI given by the absolute value of Eq.~\eqref{eq:Gamma_spt}, i.e., $\widetilde{E}_y = A \exp{\left(3/2^{2/3} \Gamma_{\rm OTSI} \,\xi^{1/3} s^{2/3} \right)}$. The prefactor $A$ is adjusted to fit the simulation data before reaching saturation.
According to the OTSI theory \cite{PSMC_2022}, this prefactor is not a constant but depends weakly on $s$ and $\xi$, which explains the slight variations in $A$ across the curves (see the legend of Fig. \ref{fig:gr_flattop}). Moreover, the analytical expression is derived using the slowly envelope approximation (invalid at positions $\xi \lesssim \lambda_p$) and a large-$s$ asymptotic expansion: these assumptions may account for the discrepancies between the simulated and theoretical curves at small $\xi$ and $s$ values. Nevertheless, the theory captures the simulation trends fairly well during the linear instability phase, as demonstrated in \cite{PSMC_2022}.


\section{OTSI simulation with the quasistatic PIC code QuaSSis}
\label{Sect2}

This section is divided into three parts. First, we recall the established equations of the QS-PIC approach \cite{Mora_1997, QuickPIC_2006, HiPACE_2014, WAND_2017}. Second, we present QuaSSis, our new QS-PIC code, which offers a major advance: the implementation of periodic transverse boundary conditions, in addition to the Dirichlet conditions used in existing codes \cite{Mora_1997, QuickPIC_2006, HiPACE_2014, WAND_2017}. This novel feature allows us to simulate transversely infinite beams. Third, we benchmark QuaSSis simulations of the OTSI against standard PIC results and theoretical predictions, examining both finite and infinite transverse beam scenarios.

\subsection{Quasistatic equations}

We adopt dimensionless units hereafter. Time is normalized by $\omega_p^{-1}$, lengths by $k_p^{-1}$, velocities by $c$, momenta by $m_e c$, charges by $e$, masses by $m_e$, densities by the initial plasma density $n_p$, scalar potentials by $m_e c^2/e$ and vector potentials by $m_ec/e$. The interaction time is considered short enough to neglect ion motion, so we assume a fixed ion background.

The QS-PIC technique, initially developed for LFWAs and PWFAs in Ref.~\cite{Mora_1997}, combines the quasistatic approximation with Hamiltonian properties for plasma electrons, yielding the $\xi$-invariant \cite{Chen_1987, Sprangle_1990, Mora_1997}
\begin{equation}
    C = \gamma - p_x - \psi \,.
\end{equation}
Here, $\gamma$ is the Lorentz factor, $p_x =\gamma v_x $ is the longitudinal momentum, $v_x$ is the longitudinal velocity and
\begin{equation}
    \psi = \varphi - A_x
\end{equation}
is a pseudo-potential with $\varphi$ and $A_x$ being the scalar and longitudinal vector potentials, respectively. Evaluating this invariant using the initial state of the plasma electrons gives
\begin{equation}
    \gamma - p_x - \psi = \gamma(\xi=0) - p_x(\xi=0) \equiv C_0 \,,
    \label{eq:qsa_invariant}
\end{equation}
and hence
\begin{equation}
    1-v_x = \frac{C_0+\psi}{\gamma} \,.
    \label{eq:hamilton_qsa}
\end{equation}
Previously developed quasistatic codes typically assume an initially cold plasma, for which $p_x(\xi=0)=0$ and $\gamma(\xi=0)=1$, resulting in $C_0=1$ \cite{qs_temp_2015, qsa_temp_2023}.

The equations of motion of the beam particles can be rewritten in terms of slow-varying $s$-derivatives noting that $d/dt = v_x d/ds$. This yields (the subscript $\perp$ referring to the $y$--$z$ transverse plane):
\begin{align}
    \frac{d \pmb{r}_\perp}{ds} &= \frac{\pmb{v}_\perp}{v_x} \,,
    \label{eq:mvt_beam_q} \\
    \frac{d \xi}{ds} &= \frac{1}{v_x} - 1 \,, \\
    \frac{d \pmb{p}}{ds} &= \frac{q_b}{v_x}\left(\pmb{E} + \pmb{v} \wedge\pmb{B} \right) \,.
    \label{eq:mvt_beam_p} 
\end{align}
Here $\pmb{r}_\perp$ is the transverse position of each particle , $\pmb{v}_\perp$ its transverse velocity, $q_b$ its charge ($q_b=-1$ for electrons and $q_b=+1$ for positrons), $\pmb{E}$ the electric field and $\pmb{B}$ the magnetic field. 
For the ultrarelativistic beam considered here, it is accurate enough to set $v_x\simeq 1$ in Eqs.~\eqref{eq:mvt_beam_q} and \eqref{eq:mvt_beam_p}.

For the plasma particles, by contrast, their equations of motion can be conveniently recast in terms of the fast-varying $\xi$-derivatives, using $d/dt = (1-v_x) d/d\xi$:
\begin{align}
    \frac{d \pmb{r}_\perp}{d \xi} &= \frac{1 }{1-v_x} \pmb{v}_\perp \,,
    \label{eq:mvt_plasma_q} \\
    \frac{d \pmb{p}}{d \xi} &=  -\frac{1 }{1-v_x} \left(\pmb{E} + \pmb{v} \wedge \pmb{B} \right) \,.
    \label{eq:mvt_plasma_p}
\end{align}

The above equations are to be supplemented with the quasistatic field equations \cite{Mora_1997, QuickPIC_2006, WAND_2017, WANPIC_2022}:
\begin{align}  
    &\Delta_\perp \psi = -\rho_p + J_{p,x} \,,
    \label{eq:psi} \\  
    &\Delta_\perp E_x =  \pmb{\nabla}_\perp \cdot \pmb{J}_\perp \,,
    \label{Ex} \\
    &\Delta_\perp B_x = -  \pmb{e}_x \cdot \left( \pmb{\nabla} \wedge \pmb{J}_\perp \right) \,,
    \label{eq:Bx} \\
    &\Delta_\perp \pmb{B}_\perp + K \pmb{B}_\perp =  \pmb{S}_\perp \,,
    \label{eq:Bperp} \\
    &E_y - B_z = - \frac{\partial \psi}{\partial y} \,,
    \label{eq:EymBz} \\
    &E_z + B_y = - \frac{\partial \psi}{\partial z} \,,
    \label{eq:EzpBy}
\end{align}
where $\pmb{J}$ is the total (plasma and beam) current density. The right-hand side of Eq.~\eqref{eq:psi} only contains the plasma sources $\rho_p$ and $J_{p,x}$ because the corresponding beam quantities cancel out when $v_x \to 1$. 
In practice, neglecting this vanishing beam contribution in Eq.~\eqref{eq:psi} allows the numerical noise to be greatly reduced in QS-PIC simulations. 

The source terms $K$ and $\pmb{S}_\perp$ involved in Eq.~\eqref{eq:Bperp} were derived in Refs.~\cite{WAND_2017, WANPIC_2022} after introducing the following distribution function (generalized here for finite-temperature plasmas through a distribution in the constant of motion $C_0$):
\begin{equation}
    f_e(t, \pmb{r}, \pmb{p}) = f_e^*(\xi,\pmb{r}_\perp,\pmb{p}_\perp,C_0) \delta \left( \gamma - p_x - \psi - C_0 \right) \,.
    \label{eq:reduced_edf}
\end{equation}
Here $\delta$ denotes the Dirac function and $f_e^*$ represents the reduced distribution function, from which we define the averaging operator $\langle . \rangle$ as
\begin{equation}
     \left< X \right> = \iiint X f_e^* \pmb{dp}_\perp dC_0 \,.
    \label{eq:averaging}
\end{equation}
Using these definitions, $K$ and $\pmb{S}_\perp$ are given by
\begin{align}
    K &= - \left< \frac{1}{\gamma-p_x}\right> \,, \label{eq:K} \\
    S_y &= \left< \frac{\gamma}{(\gamma-p_x)^2}\right> \frac{\partial \psi}{\partial z} + \left<   \frac{ p_y }{(\gamma-p_x)^2}\right>B_x \nonumber \\
    & -\left<\frac{p_z}{(\gamma-p_x)^2}\right> E_x -  \left<\frac{p_y p_z}{(\gamma-p_x)^3}\right> \frac{\partial \psi }{\partial y} -  \left< \frac{p_z^2}{(\gamma-p_x)^3}\right> \frac{\partial \psi}{\partial z} \nonumber \\
    & -\frac{\partial}{\partial y} \left(  \left<\frac{p_y p_z}{(\gamma-p_x)^2}\right>  \right) -\frac{\partial}{\partial z }\left(   \left< \frac{p_z^2}{(\gamma-p_x)^2}\right>  \right) -\frac{\partial J_{b_z}}{\partial \xi} -\frac{\partial J_x}{\partial z} \,,
    \label{eq:Sy} \\
    S_z &= -  \left<\frac{\gamma}{(\gamma-p_x)^2}\right> \frac{\partial \psi}{\partial y} +  \left<  \frac{ p_z }{(\gamma-p_x)^2}\right>B_x \nonumber \\
    &+ \left<\frac{p_y}{(\gamma-p_x)^2}\right> E_x +  \left< \frac{p_y^2}{(\gamma-p_x)^3}\right> \frac{\partial \psi }{\partial y} +  \left<\frac{p_z p_y}{(\gamma-p_x)^3}\right> \frac{\partial \psi}{\partial z} \nonumber \\
    & +\frac{\partial}{\partial y} \left(  \left< \frac{p_y^2}{(\gamma-p_x)^2}\right>  \right) +\frac{\partial}{\partial z }\left(   \left< \frac{p_y p_z}{(\gamma-p_x)^2}\right>  \right) +\frac{\partial J_{b_y}}{\partial \xi} +\frac{\partial J_x}{\partial y} \,,
    \label{eq:Sz}
\end{align}
where $\pmb{J}_b$ is the beam curent density.


For an initially cold plasma, one has $\gamma-p_x = 1 +\psi$ so that the above system simplifies to \cite{WAND_2017, WANPIC_2022}
\begin{align}
    K &= - \frac{1}{1+\psi} \,, 
    \label{eq:K_wand} \\
    S_y &= \frac{ 1}{(1+\psi)^2}  \left[ 
    \left<  \gamma \right> \frac{\partial \psi}{\partial z}  
    + \left< p_y \right> B_x
    - \left<p_z\right> E_x 
    -  \frac{\left< p_y p_z \right>}{(1+\psi)} \frac{\partial \psi }{\partial y} 
    - \frac{\left< p_z^2 \right>}{(1+\psi)} \frac{\partial \psi }{\partial z} 
    \right] \nonumber \\
    & -\frac{\partial}{\partial y} \left(   \frac{ \left<p_y p_z \right>}{(1 + \psi)^2}  \right) -\frac{\partial}{\partial z }\left(    \frac{\left<p_z^2 \right> }{(1 + \psi)^2} \right) -\frac{\partial J_{b_z}}{\partial \xi} -\frac{\partial J_x}{\partial z} \,,
    \label{eq:Sy_wand} \\
    S_z &= \frac{ 1}{(1+\psi)^2}  \left[ 
    - \left< \gamma \right>  \frac{\partial \psi}{\partial y}  
    + \left< p_z \right> B_x
    + \left<p_y\right> E_x 
    + \frac{\left<  p_y^2 \right> }{(1+\psi)}\frac{\partial \psi }{\partial y} 
    + \frac{\left<  p_y p_z \right>}{(1+\psi)} \frac{\partial \psi }{\partial z} 
    \right] \nonumber \\
    & +\frac{\partial}{\partial y} \left(  \frac{\left<p_y^2 \right>}{(1 + \psi)^2}  \right) +\frac{\partial}{\partial z }\left(   \frac{\left<p_y p_z \right> }{(1 + \psi)^2} \right) +\frac{\partial J_{b_y}}{\partial \xi} +\frac{\partial J_x}{\partial y} \,.
    \label{eq:Sz_wand}
\end{align}

\subsection{Numerical scheme}

The QS-PIC algorithm follows the same approach as standard PIC codes \cite{QuickPIC_2006, WAND_2017}, except that the dynamics of the beam and the plasma are decoupled. The computation therefore proceeds in two steps.
In the first step, the plasma and field quantities are advanced in $\xi$ through Eqs.~\eqref{eq:mvt_plasma_q}--\eqref{eq:Bperp}. During this step, the beam density and current are held constant, depending only on $\xi$ and $\pmb{r}_\perp$. In the second step, the positions and momenta of the beam MPs are advanced in $s$ through Eqs.~\eqref{eq:mvt_beam_q} and \eqref{eq:mvt_beam_p}, where the fields are assumed to depend only on $\xi$ and $\pmb{r}_\perp$. The spatial step $\Delta s$ is chosen large enough to greatly reduce the simulation time significantly, while being small enough to accurately capture the beam dynamics.

This two-step process is repeated iteratively. After each update of the beam over the propagation distance $\Delta s$, the quasistatic plasma and field equations are solved again. This cycle continues until the end of the interaction. The computation is thus decomposed into two loops: one for the beam evolution ($s$ loop) and another for the plasma and field evolution ($\xi$ loop). In the latter, $\xi$ serves as the new ``time variable'', reducing the dimensionality of the problem and requiring only a 2D ($y,z$) mesh.


The QS-PIC scheme can be detailed as follows:
\newline

\noindent 
\textit{
\textbf{Beginning of the $s$ loop}
\begin{itemize}
    \item Deposit the charge and current densities of the beam MPs at each cell of the ($\xi, y, z$) mesh, as in a standard PIC code.
    \item \textbf{Beginning of the $\xi$ loop}
    \begin{itemize}
        \item Deposit the charge and current densities $\rho_p$ and $\pmb{J}_p$ of the plasma MPs together with the source terms $K$ and $\pmb{S}_\perp$ on the nodes of the quasistatic $(y,z)$ mesh.
        \item Update the fields using Eqs.~\eqref{eq:psi}--\eqref{eq:Bperp}. 
        \item Interpolate the fields from the $(y,z)$ mesh nodes to the plasma MPs.
        \item Update the plasma MPs using Eqs.~\eqref{eq:mvt_plasma_q} and \eqref{eq:mvt_plasma_p}.
    \end{itemize}
    \item \textbf{End of the $\xi$ loop}
    \item Interpolate the fields from the ($\xi,y,z$) mesh to the beam MPs.
    \item Update the beam MPs using Eqs.~\eqref{eq:mvt_beam_q} and \eqref{eq:mvt_beam_p}.
\end{itemize}
\textbf{End of the $s$ loop}
}
\newline

We have implemented this scheme into QuaSSis, a 2D3V Python code optimized with Numba, dedicated to the study of relativistic beam-plasma instabilities. Due to the 2D spatial geometry, only the coordinates $\xi$ and $y$ are considered. A Cartesian mesh is used with the transverse $y$-axis discretized into $N_y+1$ equally spaced nodes. QuaSSis can handle plasmas with finite initial temperatures through Eqs.~\eqref{eq:K}--\eqref{eq:Sz}.

The deposition of plasma quantities in the $\xi$ loop follows the standard QS-PIC approach \cite{QuickPIC_2006, HiPACE_2014}. The plasma charge ($\rho_{p,k}$) and current ($\pmb{J}_{p,k}$) densities at the transverse position $y_k$ of the $k^{\rm th}$ node are given by
\begin{align}
    \rho_{p,k} &= \rho_I -\frac{1}{V} \sum_i \frac{w_i}{1-v_{x,i}} \mathcal{S}(y_i,y_k) \,,
    \label{eq:depot_rho} \\
    \pmb{J}_{p,k} &= -\frac{1}{V} \sum_i \frac{w_i}{1-v_{x,i}}\pmb{v}_i \mathcal{S}(y_i, y_k) \,,
    \label{eq:depot_J} 
\end{align}
where $\rho_I$ is the constant ion charge density, $w_i$ and $y_i$ are the numerical weight and tranverse position of MP $i$, $V$ is the ``volume'' of the cell, and $\mathcal{S}$ is the interpolation (shape) function \cite{Birdsall_2004}. In both equations, $\psi$ is determined at the MPs' positions through Eq.~\eqref{eq:hamilton_qsa}, not from mesh interpolation. QuaSSis handles shape functions from 0th to 3rd order, but all simulations in this paper use a 3rd-order shape function.

The factor $w_i/(1-v_{x,i}) =w_i \gamma_i/(C_{0,i}+\psi_i)$ in the above equations can be viewed as a quasistatic-corrected MP weight. More generally, one can define the discrete average of the particle quantity $X$ at node $k$ as 
\begin{equation}
    \left< X \right>_k= \frac{1}{V} \sum_i w_i X_i \mathcal{S}(y_i,y_k) \,.
    \label{eq:discrete_averaging}
\end{equation}
Using this definition, one has $\rho_{p,k} = -\left< \frac{\gamma}{C_{0} + \psi} \right>_k = -\left< \frac{1}{1-v_x} \right>_k $, $\pmb{J}_{p,k} = -\left<\frac{\pmb{v}}{1-v_x} \right>_k $, and $K_k = \left< \frac{1}{\gamma - p_x} \right>_k$. The above operator also enables one to convert the continuous expressions of $S_y$ and $S_z$,  Eqs.~\eqref{eq:Sy} and \eqref{eq:Sz}, into their discretized variants. In QuaSSis, if the plasma is initialized with no initial temperature ($C_0=1$), the general Eqs.~\eqref{eq:K}--\eqref{eq:Sz} are still used, although a faster and simpler alternative would be to directly compute Eqs.~\eqref{eq:K_wand}--\eqref{eq:Sz_wand}.

The field equations \eqref{eq:psi}-\eqref{eq:Bperp} are solved using a finite-difference method as in Ref.~\cite{LCode_1999} for Cartesian geometry or Ref.~\cite{QPAD_2021} for cylindrical geometry. 

To discretize these equations, we use a constant step $\Delta \xi$ and define $X_k^n \equiv X(\xi = n \Delta \xi \equiv \xi^n, y = k\Delta y\equiv y_k)$, with $k \in \left[0, N_y\right]$. Under homogeneous Dirichlet boundary conditions, the set of Eqs.~\eqref{eq:psi}--\eqref{eq:Bperp} is solved by inverting the discrete system
\begin{align}
    &\psi_0^n = \psi_{N_y}^n = 0 \,,
    \label{eq:df_CL_psi} \\
    &\forall k \in \left[1,N_y-1\right] \quad \frac{\psi_{k+1}^n -2\psi_k^n + \psi_{k-1}^n}{\Delta y^2} = -\rho_k^n + J_{x,k}^n  \,,
    \label{eq:df_mat_psi} \\
    &E_{x,0}^{n} = E_{x,N_y}^n = 0 \,,
    \label{eq:df_CL_Ex} \\
    &\forall k \in \left[1,N_y-1\right] \quad \frac{E_{x,k+1}^n -2E_{x,k}^n + E_{x,k-1}^n}{\Delta y^2 } = \frac{J_{y,k+1}^n - J_{y, k-1}^n}{2 \Delta y} \,, 
    \label{eq:df_mat_Ex} \\
    &B_{x,0}^{n} = B_{x,N_y}^n = 0 \,,
    \label{eq:df_CL_Bx} \\
    &\forall k \in \left[1,N_y-1\right] \quad \frac{B_{x,k+1}^n -2B_{x,k}^n + B_{x, {k-1}}^n}{\Delta y^2} = -\frac{J_{z,k+1}^n - J_{z,k-1}^n}{2 \Delta y} \,, 
    \label{eq:df_mat_Bx} \\
    &\pmb{B}_{\perp,0}^n = \pmb{B}_{\perp, N_y}^n = \pmb{0} \,,
    \label{eq:df_CL_Bperp} \\
    &\forall k \in \left[1,N_y-1\right] \quad 
            \frac{1}{\Delta y^2} \pmb{B}_{\perp,k+1}^n + \left(K_k^n -\frac{2}{\Delta y^2} \right) \pmb{B}_{\perp,k}^n + \frac{1}{\Delta y^2} \pmb{B}_{\perp,k-1}^n
            = \pmb{S}_{\perp,k}^n \,,
    \label{eq:df_mat_Bperp} \\
    &E_{y,0}^n = E_{y, N_y}^{n} = 0 \,,
    \label{eq:df_CL_Ey} \\
    &\forall k \in \left[1,N_y-1\right] \quad E_{y,k}^n =  B_{z,k}^n - \frac{\psi_{k+1}^n - \psi_{k-1}^n}{2\Delta y} \,,   
    \label{eq:df_mat_Ey} \\
    &\forall k \in \left[0,N_y\right] \quad E_{z,k}^n = -B_{y,k}^n  \,.
    \label{eq:df_mat_Ez}
\end{align}
The fields \eqref{eq:df_CL_psi}--\eqref{eq:df_mat_Ez} are solved in matrix form using the lower-upper (LU) decomposition from the LAPACK library \cite{Lapack1999}. In Eq.~\eqref{eq:df_mat_Bperp}, $\pmb{S_\perp}_k^n$ is evaluated from the 2D form of Eqs.~\eqref{eq:Sy} and \eqref{eq:Sz}. To simplify the notation, the operator \eqref{eq:discrete_averaging} is given an exponent $n$ to indicate that the distribution of the plasma MPs is considered at the quasistatic time $\xi^n$:
\begin{align}
    S_{y,k}^n &= 
    \left< \frac{p_y}{(\gamma-p_x)^2}\right>_k^n B_{x, k}^n 
    - \left<\frac{p_z}{(\gamma-p_x)^2}\right>_k^n E_{x,k}^n 
    -  \left<\frac{p_y p_z}{(\gamma-p_x)^3}\right>_k^n \frac{\psi_{k+1}^n - \psi_{k-1}^n}{2 \Delta y} \nonumber \\
    &-\frac{1}{2 \Delta y} \left(  \left<\frac{p_y p_z}{(C_0 + \psi)^2}\right>_{k+1}^n -   \left<\frac{p_y p_z}{(C_0 + \psi)^2}\right>_{k-1}^n \right) 
    -\frac{1 }{2 \Delta \xi} \left( J_{b_z,k}^{n+1} -  J_{b_z,k}^{n-1} \right)  \,,
    \label{eq:Sy_discret} \\
    S_{z,k}^n &= 
    \left< \frac{ p_z }{(\gamma-p_x)^2}\right>_k^n B_{x,k}^n
    + \left<\frac{p_y}{(\gamma-p_x)^2}\right>_k^n E_{x,k}^n  
    + \left( \left<\frac{p_y^2}{(\gamma-p_x)^3}\right>_k^n  - \left<\frac{\gamma}{(\gamma-p_x)^2}\right>^n \right) \frac{\psi_{k+1}^n - \psi_{k-1}^n}{2 \Delta y} \nonumber \\
    &+\frac{1}{2 \Delta y} \left( \left<\frac{p_y^2}{(C_0 + \psi)^2}\right>_{k+1}^n  - \left<\frac{p_y^2}{(C_0 + \psi)^2}\right>_{k-1}^n + J_{x,k+1}^n - J_{x,k-1}^n\right)
    +\frac{1 }{2 \Delta \xi} \left( J_{b_y,k}^{n+1} -  J_{b_y,k}^{n-1} \right) \,,
    \label{eq:Sz_discret}
\end{align}
where $J_{b_y,k}^{n-1}$, $J_{b_y,k}^{n+1}$, $J_{b_z,k}^{n-1}$, $J_{b_z,k}^{n+1}$, $J_{b_x,k-1}^n$, $J_{b_x,k+1}^n$ refer to the beam current components known on the ($\xi,y$) mesh at the start of the $s$ loop.

Field interpolation is performed as in the standard PIC method, i.e., the fields $\pmb{E}_i$ and $\pmb{B}_i$ seen by MP $i$ are  computed as
\begin{align}
    \pmb{E}_i &= \sum_k \pmb{E}_k \mathcal{S}(y_i, y_k) \,, \\
    \pmb{B}_i &= \sum_k \pmb{B}_k \mathcal{S}(y_i, y_k) \,,
\end{align}
using the same shape function as in the deposition step to avoid self-forces \cite{Birdsall_2004}.

The equations of motion for the plasma MPs, Eqs.~\eqref{eq:mvt_plasma_q} and \eqref{eq:mvt_plasma_p}, are solved using the 5th-order Adams-Bashforth scheme \cite{butcher2016}, as is done in the HiPACE code \cite{HiPACE_2014, HiPACE_2022}. In detail, for a particle quantity $X \in \{y, p_x, p_y, p_z\}$, we define $F^n \equiv \left(dX/d\xi \right)(\xi=\xi^n)$ and advance $X^n \equiv X(\xi = \xi^n)$ via
\begin{equation}
    X^{n+1} = X^n + \Delta\xi \sum_{j=0}^{4} \alpha_j F^{n-j} \,,
    \label{eq:AB5}
\end{equation}
where the coefficients $\{\alpha_j\}_{0\le j \le 4}$ are given by \cite{butcher2016}:
\begin{equation}
    \alpha_0 = \frac{1901}{720}, \quad \alpha_1 = -\frac{1387}{360}, \quad \alpha_2 = \frac{109}{30}, \quad \alpha_3 = -\frac{637}{360}, \quad \alpha_4 = \frac{251}{720} \,.
\end{equation}

\subsection{Test simulation of OTSI with a finite-size beam}

To validate its numerical implementation, QuaSSis is first benchmarked against the PIC code CALDER. The example presented here involves a Gaussian-distributed beam that could be delivered at the SLAC/FACET II facility \cite{Yakimenko_2019, PSMC_2022}. The beam is characterized by an energy of $10\,\rm GeV$, a charge of $2\,\rm nC$, a longitudinal root-mean-square (rms) size of $\sigma_x = 5\,\rm \mu m$, a transverse rms size of $\sigma_y = 10\,\rm \mu m$, a transverse normalized emittance of $\epsilon = 3\, \rm mm \, mrad$, and a peak density of $n_b = 1.5 \times 10^{18}\,\rm cm^{-3}$. It is injected into a uniform cold ($T_p = 0.01 \, \rm eV $) plasma with $n_p = 5 \times 10^{19}\,\rm cm^{-3}$ density, which could be created through the ionization of a gas target. Both simulations are run with $1200$ cells in the longitudinal direction and $2400$ cells in the transverse direction. In CALDER, $\Delta x = \Delta y = 0.5 \,k_p^{-1}$ ($0.038\,\rm \mu m$) and $\Delta t = 0.35\,\rm \omega_p^{-1}$ ($0.088\,\rm fs$). In QuaSSis, $\Delta \xi = \Delta y = 0.5\, k_p^{-1}$ ($0.038 \, \rm \mu m$) and $\Delta s = 100\,k_p^{-1}$ ($\Delta s / c = 502.9 \,\rm fs$).
The plasma and beam electron populations are initially represented by 10 regularly spaced MPs per cell. Their numerical weight is randomly drawn as detailed in Sec.~\ref{sect_noise} to control the initial noise in the simulation. In QuaSSis, the beam is initialized in the plasma, while in CALDER, it is initialized outside the plasma and propagated ballistically until the plasma response becomes quasistatic. The two simulations are synchronized so that QuaSSis starts when the CALDER beam stops being ballistic.

\begin{figure*}
    \centering
    \includegraphics[scale=0.35]{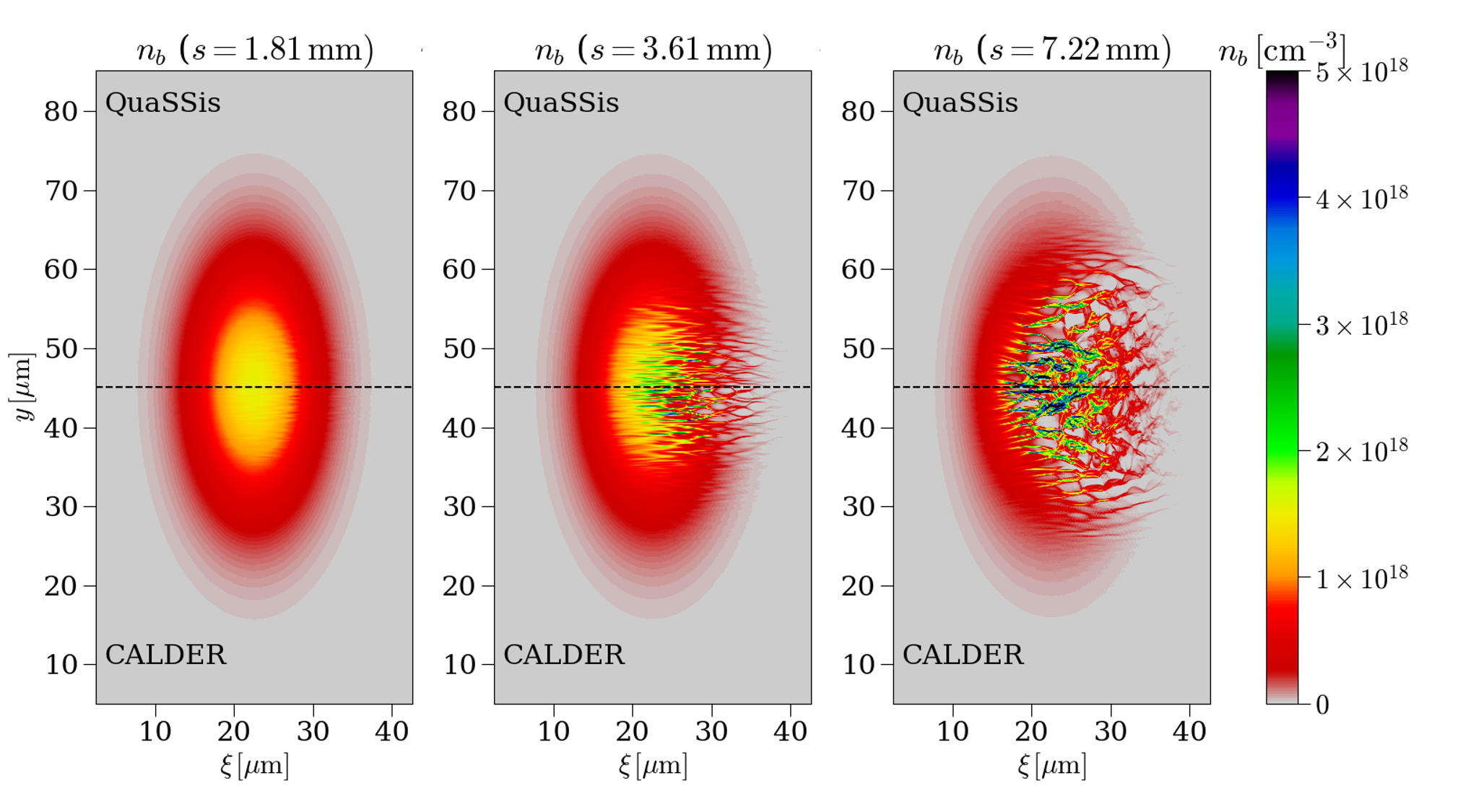}
    \caption{Density map of the beam electrons from the standard PIC (CALDER) simulation (bottom half-plane) and from the QS-PIC (QuaSSis) simulation (top half-plane) at three propagation distances ($s$). Both simulations consider a $10\,\rm GeV$ beam with $n_b = 1.5 \times 10^{18}\,\rm cm^{-3}$ peak density propagating in a uniform plasma with $n_p = 1 \times 10^{19}\,\rm cm^{-3}$ density. The initial beam profile is Gaussian in both longitudinal and transverse directions, with respective rms sizes $\sigma_x = 5\,\rm \mu m$ and $\sigma_y = 10 \,\rm \mu m$.}
    \label{fig:SMC_Quassis}
\end{figure*}

The CALDER and QuaSSis results are compared in Fig.~\ref{fig:SMC_Quassis}. The three panels display the beam density profile at different propagation distances, with the top and bottom half-planes corresponding to QuaSSis and CALDER, respectively. The two simulations closely agree in terms of localization, shape, and amplitude of the beam modulations induced by the OTSI. Early in the interaction ($s = 1.81\,\rm mm$, left panel), faint modulations begin to emerge at the trailing edge of the beam. By $s = 3.61\,\rm mm$ (middle panel), they have become much stronger, forming a chevron-type pattern that extends from the center to the rear of the beam, while the front and transverse edges remain undisturbed. Finally, at $s = 7.22\,\rm mm$ (right panel), the OTSI is fully developed throughout most of the beam, resulting in highly compressed filaments (at about three times their initial density) and completely beam-depleted regions.

Not only is QuaSSis capable of reproducing accurately the CALDER results, it does so at a significantly lower computational cost. Specifically, the Fortran CALDER code required 2.07~h of wall-clock time on 2500 cores (utilizing hybrid MPI and OpenMP parallelism), totaling 5170 core.hours. In contrast, the Python QuaSSis code, partially optimized with Numba, completed the same simulation on a single core in 8.2~h only, demonstrating a very substantial reduction in computational resources. These first results highlight the potential of a quasistatic code like QuaSSis as a highly accurate and computationally efficient tool for studying relativistic beam-plasma instabilities.

\subsection{Transverse periodic boundary conditions}

To replicate the CALDER simulation presented in Fig.~\ref{fig:CLP_Calder} using QuaSSis and compare it with the spatiotemporal model of Ref.~\cite{PSMC_2022}, periodic boundary conditions are needed. Implementing these conditions is straightforward for MPs (both plasma and beam): any particle exiting the simulation domain through a transverse boundary is reintroduced on the opposite side, conserving its energy and momentum. However, handling periodic boundary conditions for field resolution is more complex. These conditions impose that the field values and their transverse derivatives match on opposite sides of the domain. Consequently, our previous approach for solving the Poisson equations \eqref{eq:psi}--\eqref{eq:Bx} is no longer viable because the resulting matrix in Cartesian coordinates becomes singular (non-invertible). When using an FFT solver, the mean field value remains undetermined, becoming a free parameter. A solution to this issue is given in Ref.~\cite{Birdsall_2004}, but we choose to stay within the framework of finite differences.


Instead, to solve the Poisson equations governing the fields $\psi$, $E_x$ and $B_x$, we first predict their boundary values (i.e., at $k=0$ and $k=N_y$) based on the field information from the previous iteration, $\xi^{n-1}$. These boundary values are then used as Dirichlet boundary conditions, consistent with our previous approach. The spatiotemporal equations yielding $\psi$, $E_x$ and $B_x$ at the transverse boundaries are
\begin{align}  
    &\frac{\partial \psi}{\partial \xi} = E_x \,,
    \label{eq:psi_ksi} \\
    &\frac{\partial E_x}{\partial \xi} = \frac{\partial B_z}{\partial y} - \frac{\partial B_y}{\partial z} - J_x \,,
    \label{eq:Ex_ksi} \\
    &\frac{\partial B_x}{\partial \xi} = \frac{\partial E_z}{\partial y} - \frac{\partial E_y}{\partial z} \,.
    \label{eq:Bx_ksi}
\end{align} 
These equations originate from the definition of the pseudo-potential $\psi$, the $x$-component of the quasistatic Maxwell-Amp\`ere equation, and the $x$-component of the quasistatic Maxwell-Faraday equation, respectively. 

We have implemented the Runge-Kutta, Adams-Bashforth, and Adams-Moulton methods \cite{butcher2016} to solve the above set of equations. The Runge-Kutta and Adams-Moulton methods rely on an explicit predictor-corrector scheme, involving a more complex and time-consuming iterative solver. The Adams-Bashforth method is simpler but known to be less stable. Although the three schemes yield similar results, we have kept the 5th-order Adams-Bashforth method, which ensures faster computation and is already used to push the plasma MPs \cite{HiPACE_2014}. Equations \eqref{eq:psi_ksi}--\eqref{eq:Bx_ksi} are then discretized as follows:
\begin{align}
        &\psi_0^n = \psi_{N_y}^{n} = \psi^{n-1}_0 + \Delta\xi \sum_{i=0}^{4} \alpha_i E_{x,0}^{n-1-i} \,,
        \label{eq:AB5_psi} \\
        &E_{x,0}^n = E_{x,N_y}^n = E_{x,0}^{n-1} + \Delta\xi \sum_{i=0}^4 \alpha_i \left(  \frac{B_{z,1}^{n-1-i} - B_{z, {N_y-1}}^{n-1-i}}{2 \Delta y}  - J_{x,0}^{n-1-i} \right) \,,
        \label{eq:AB5_Ex} \\
        &B_{x,0}^n = B_{x,N_y}^n = B_{x,0}^{n-1} + \Delta\xi \sum_{i=0}^4 \alpha_i \frac{E_{z,1}^{n-1-i} - E_{z,N_y-1}^{n-1-i}}{2 \Delta y } \,. 
        \label{eq:AB5_Bx}
\end{align}
Equations~\eqref{eq:df_mat_psi}, \eqref{eq:df_mat_Ex} and \eqref{eq:df_mat_Bx} are then used to compute $\psi$, $E_x$ and $B_x$ across the full domain. 

For the $\pmb{E}_\perp$ and $\pmb{B}_\perp$ fields, the periodic boundary condition is applied by inverting the following discretized system: 
\begin{align}
    &\frac{1}{\Delta y^2} \pmb{B}_{\perp,1}^n + \left( \frac{-2}{\Delta y^2} + K_0^n \right) \pmb{B}_{\perp,0}^n + \frac{1}{\Delta y^2} \pmb{B}_{\perp,N_y-1}^n = \pmb{S}_{\perp,0}^n \,,
    \label{eq:df_CLP_Bperp0} \\
    &\pmb{B}_{\perp,N_y}^n = \pmb{B}_{\perp,0}^n \,,
        \label{eq:df_CLP_BperpN} \\
    &\forall k \in \left[1,N_y-1\right]
    \quad \frac{1}{\Delta y^2} \pmb{B}_{\perp,k+1}^n + \left( \frac{-2}{\Delta y^2} + K_k^n \right) \pmb{B}_{\perp,k}^n + \frac{1}{\Delta y^2 } \pmb{B}_{\perp,k-1}^n = \pmb{S}_{\perp,k}^n \,,
    \label{eq:df_CLP_mat_Bperp} \\
    &E_{y,0}^n = E_{y,N_y}^n = B_{z,0}^n - \frac{\psi_1^n - \psi_{N_y-1}^n}{2 \Delta y} \,,
    \label{eq:df_CLP_Ey0} \\
    &\forall k \in \left[1,N_y-1\right]
    \quad E_{y,k}^n = B_{z, k}^n - \frac{\psi_{k+1}^n - \psi_{k-1}^n}{2 \Delta y} \,,
    \label{eq:df_CLP_mat_Ey} \\
    &\forall k \in \left[0,N_y\right] \quad E_{z,k}^n = - B_{y,k}^n
    \label{eq:df_CLP_mat_Ez} \,.
\end{align}

\subsection{Simulation of OTSI with a transversely uniform beam}

As a validation test of QuaSSis with periodic boundary conditions, we reproduce the CALDER simulation presented in Sec.~\ref{Sect1}, with an electron-positron pair beam that is flat-top in the longitudinal direction and uniform in the transverse direction. The numerical parameters are those employed in CALDER (10 MPs per cell for the beam and plasma species, $\Delta \xi = \Delta y = 0.05\, k_p^{-1}$) except for $\Delta s = 150.43\, \mu \mathrm{m} = 200\,k_p^{-1}$ ($\sim 5700\times \Delta t$ used in CALDER).

\begin{figure}
    \centering
    \includegraphics[scale=0.3]{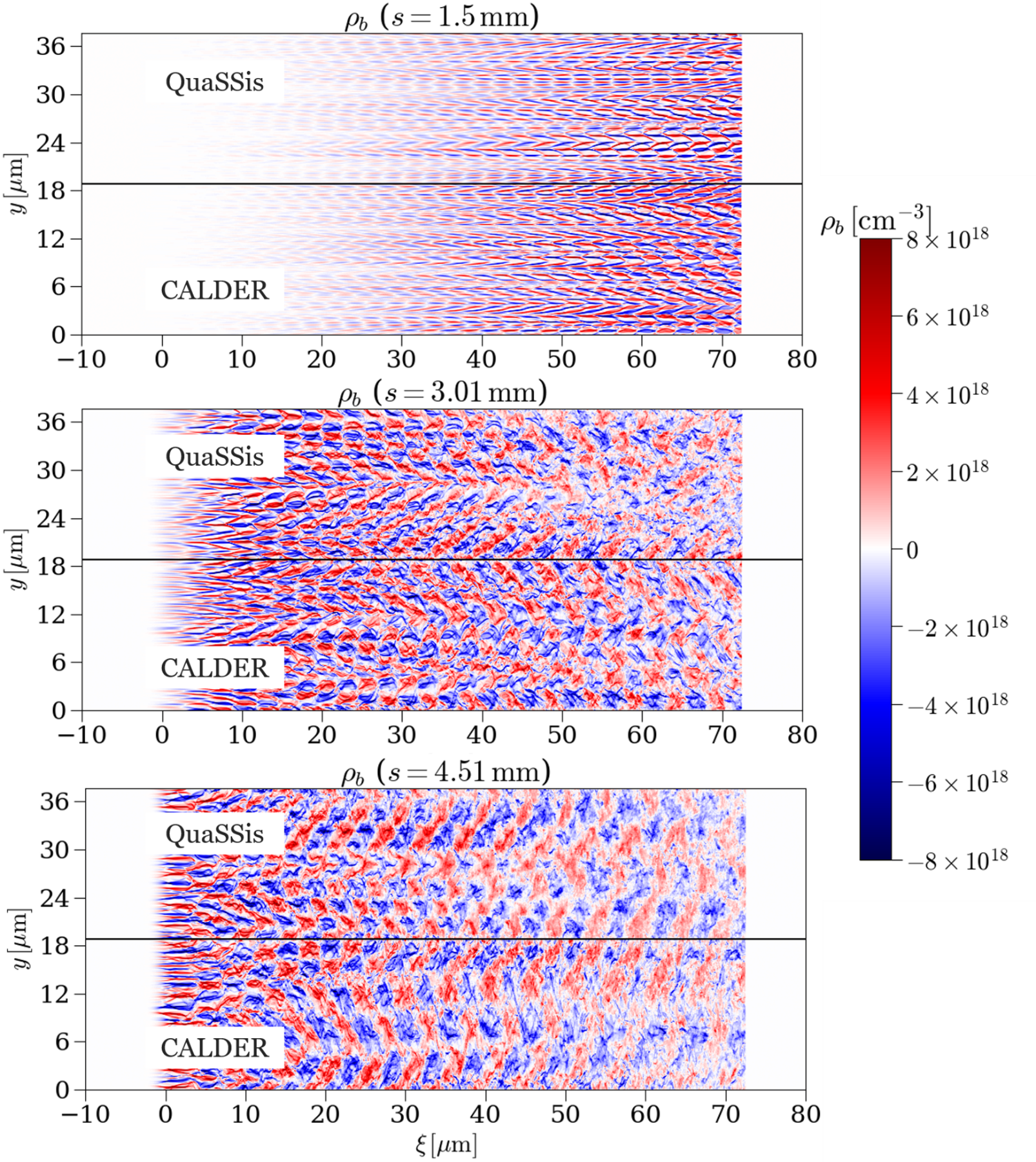} 
    \caption{Same as Fig.~\ref{fig:CLP_Calder} but with the full PIC (CALDER) results (bottom half-plane) compared to the QS-PIC (QuaSSis) results (top half-plane).
    }
\label{fig:CLP}
\end{figure}

Figure~\ref{fig:CLP} compares the beam evolution as predicted by the two codes. As in the finite-size beam case [Fig.~\ref{fig:SMC_Quassis}], we observe excellent agreement between the QuaSSis (top half-plane) and CALDER (bottom half-plane) results. The CALDER simulation took $1.3\,\rm h$ on 2500 cores, for a total of 3272~core.hours, while the QuaSSis simulation was run on a single core in just $3.1\,\rm h$. This once again demonstrates that QuaSSis can achieve computational speedups of several orders of magnitude.

\begin{figure}
    \centering
    \includegraphics[scale=0.4]{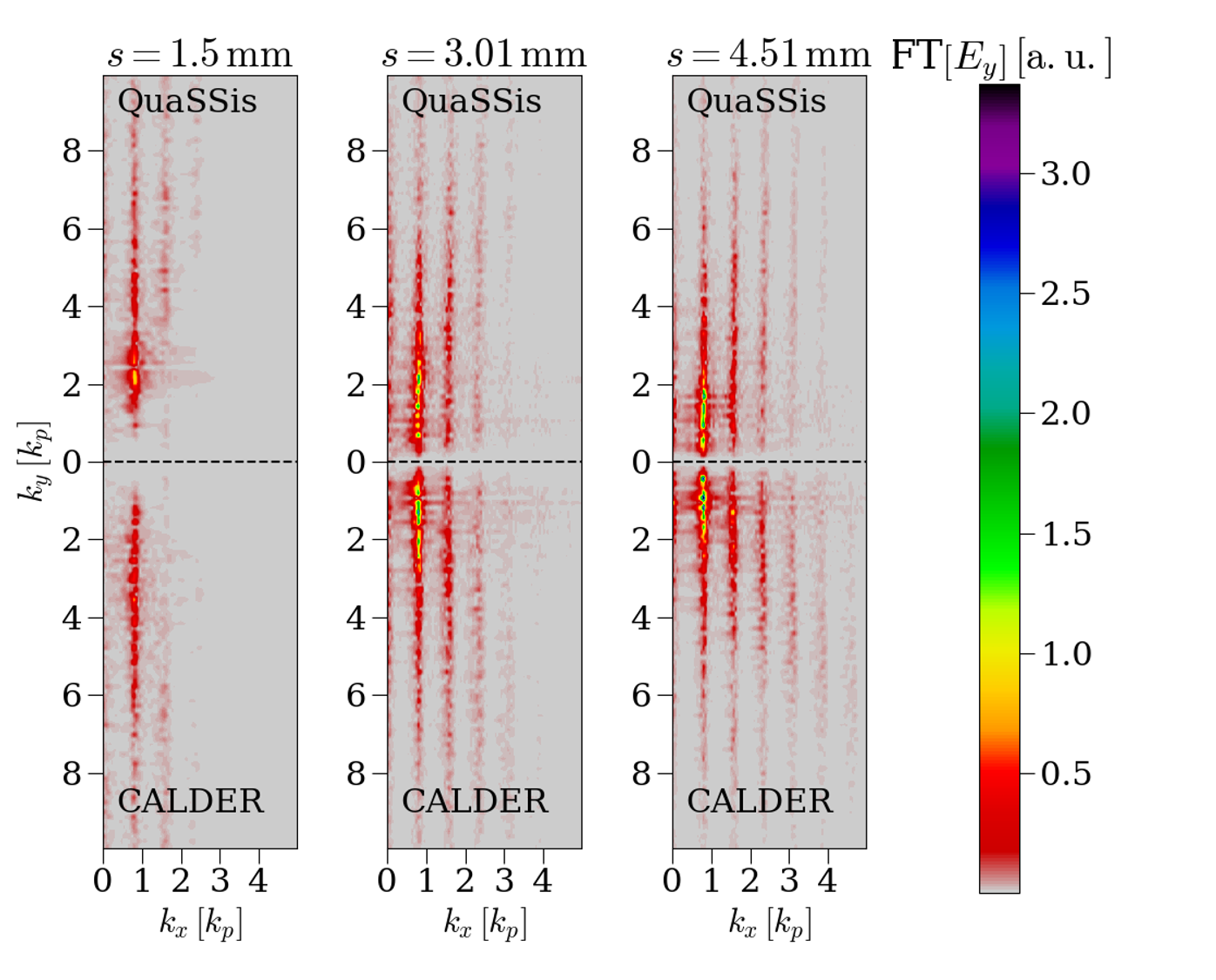} 
    \caption{Same as Fig.~\ref{fig:Fourier} but with the CALDER results (bottom half-plane) compared to the QuaSSis results (top half-plane).
    }
\label{fig:CLP_QvC}
\end{figure}

\begin{figure}
    \centering
    \includegraphics[scale=0.25]{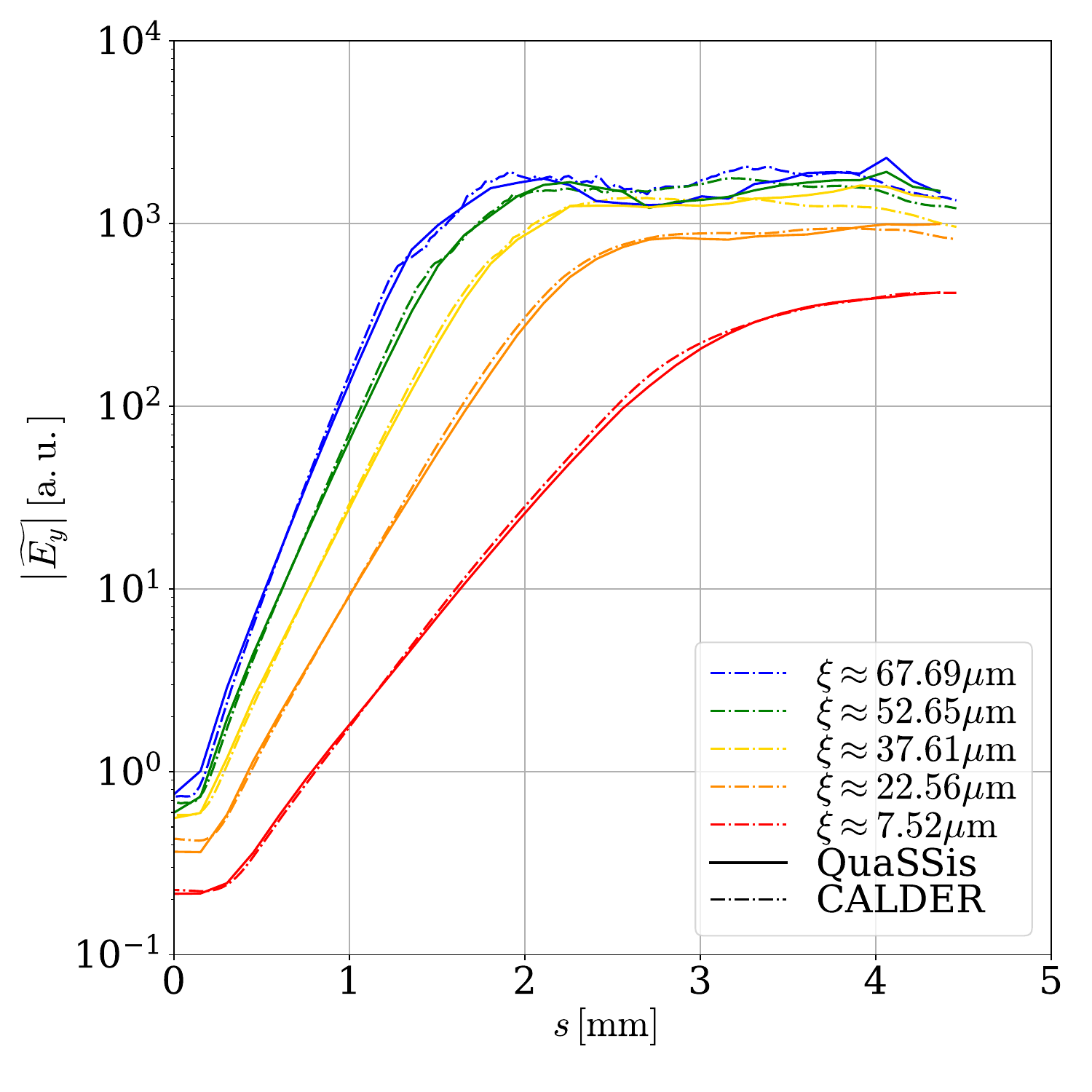}
    \caption{Spectral amplitude $\widetilde{E}_y$ [Eq.~\eqref{E_y_tilde}] as a function of the propagation distance $s$ for various positions $\xi$ in the beam. The QuaSSis results (solid lines) agree well with the CALDER results (dashed lines, already plotted in Fig.~\ref{fig:gr_flattop}). 
    }
    \label{fig:gr_QvC}
\end{figure}

Figure~\ref{fig:CLP_QvC} shows that QuaSSis also captures accurately the 2D Fourier spectra of $E_y(\xi,y)$ predicted by CALDER. The continuum of OTSI modes centered on $k_x \simeq k_p$ and its harmonic replica are reproduced with similar amplitude. The instability growth is detailed in Fig.~\ref{fig:gr_QvC}, which plots the evolution of the spectral amplitude $\left \vert \widetilde{E}_y \right \vert$ [Eq.~\eqref{E_y_tilde}] as a function of $s$ for different $\xi$, as in Fig.~\ref{fig:gr_flattop}. Again, an excellent match is found between the QuaSSis (dash-dotted) and CALDER (solid) curves, during both the linear and nonlinear phases of the instability, showing that the saturation mechanism is well captured by the QS-PIC method. Evidently, while not explicitly shown in Fig.~\ref{fig:gr_QvC}, the QuaSSis results closely align with the predictions of the spatiotemporal OTSI theory.

These successful tests indicate that QuaSSis is an efficient tool for simulating the unstable propagation of ultrarelativistic beams through background plasmas, offering a dramatic speedup in computation compared to the full PIC method. It is suitable for both realistic finite-size beam scenarios and, thanks to the periodic boundary scheme proposed here, idealized transversely uniform density conditions. This makes it valuable for benchmarking simplified instability theories.


\section{Influence of initial fluctuation level}
\label{sect_noise}

\subsection{Tuning the initial simulation noise}

Instabilities develop from initial fluctuations in the electromagnetic fields or particle charge/current densities. In PIC simulations, these fluctuations originate from the high-level noise intrinsic to finite particle count and spatiotemporal discretization. Although fluctuations related to discrete particle effects have a physical reality \cite{Ruyer_2013}, they tend to be dramatically amplified in PIC simulations compared to real, finite-temperature plasmas. This is acceptable for theoretical studies as long as the fluctuation level lies well within the perturbative regime and is orders of magnitude below the saturation level of the instabilities at play. However, to mimic realistic interaction conditions, it can be beneficial to employ a method able to easily tune the initial noise.

Several methods exist to adjust the PIC simulation noise, such as changing the resolution, the number and shape factor of MPs, using filtering schemes, or even modifying the background plasma temperature. Here, we test different MP initialization methods with fixed resolution, shape factor, plasma temperature and without filtering, to isolate the initialization effect while maintaining simulation accuracy. 

Standard PIC simulations typically model uniform density plasmas with equally weighted MPs. This results in very low initial noise if MPs are regularly spaced within cells and have zero (or very weak) initial temperature. By contrast, randomizing the MP positions often leads to significant density fluctuations, generating electromagnetic noise. For a given resolution and shape factor, reducing the noise level requires increasing the number of MPs per cell ($N_{ppc}$), but this can be computationally expensive given the $N_{ppc}^{-1/2}$ scaling of fluctuations.

As an alternative, we propose to keep the regular spacing of the MPs while modifying their weight. Specifically, for a plasma species with local number density $n$, instead of a uniform weight $w=n V/N_{ppc}$, we assign random weights to MPs according to
\begin{equation}
    w_i = w \left( 1 + \varepsilon \mathcal{U} \right) \,,
    \label{eq:w_i_random}
\end{equation}
where $\mathcal{U}$ is a uniform random number over $\left[-1,1\right]$, and $\varepsilon$ is a small $(\ll 1)$ parameter that determines the fluctuation amplitude. 

We repeated the QuaSSis simulation presented in the previous section (10~GeV beam energy, $\alpha = 0.06$, infinite transverse beam with constant density and transverse periodic conditions, $T_p = 0.01 \,\rm eV$)  but varying the MP initialization method. First, the MPs were initialized with equal weights and random positions. Next, we used regularly spaced MPs with random weights drawn according to Eq.~\eqref{eq:w_i_random}. Random weighting was applied to either plasma MPs or beam MPs (in which case the tuning parameter was renamed $\varepsilon_b$). Good agreement was found between the simulated and theoretical spatiotemporal growth rates in all cases. However, these tests confirmed that when MP positions are randomly drawn, a high MP number per cell is necessary to reduce initial noise. We also found that, for the beam-plasma parameters studied, assigning random weights to the plasma MPs alone resulted in very low noise amplitude, rendering it unsuitable for effectively tuning the noise level to a desired magnitude. Instead, assigning random weights to the beam MPs offered good control over the initial noise without modifying the computation time. Consequently, we adopted the latter approach for all QuaSSis and CALDER simulations reported previously (i.e., for a Gaussian beam with $\varepsilon_b=0.1$ in Fig.~\ref{fig:SMC_Quassis} and for a flat-top beam with $\varepsilon_b=0.05$ in Fig.~\ref{fig:CLP}).

\subsection{Noise control with random weighting}
 
We now focus on the use of regularly spaced MPs with random weighting for the beam MPs. To understand how the parameter $\varepsilon_b$ controls the initial electric fluctuations, we assume that these are driven by fluctuations in the beam's charge ($\rho_b$) and current ($\pmb{J}_b$) densities. Our starting point is then the wave equation for the electric field, considering only the beam contribution:
\begin{equation}
    \left( \Delta - \frac{\partial^2}{\partial t^2} \right) \pmb{E} = \pmb{\nabla} \rho_b + \frac{\partial \pmb{J}_b}{\partial t} \,.
    \label{eq:WE}
\end{equation}
In a 2D geometry and within the quasistatic approximation, this simplifies into
\begin{equation}
    \frac{\partial^2 E_y}{\partial y^2}  = \frac{\partial \rho_b}{\partial y} + \frac{\partial J_{b y}}{\partial \xi} \,.
    \label{eq:Helmoltz_Ey}
\end{equation}
To compute the standard deviation $\delta E_y$, we assume that the discrete values of $\rho_{b,k}^n$ and $J_{b,k}^n$ fluctuate randomly with standard deviations $\delta \rho_b$ and $\delta J_{b_y}$, respectively, with no correlation between different cells $k$ and zero mean values. This leads to
\begin{equation}
        \delta E_y \simeq \alpha \Delta y  \left( \delta \rho_b^2+ \frac{\Delta y^2}{\Delta \xi^2} \delta J_{b y}^2 \right)^{1/2} \,,
        \label{eq:Approx_Ey}
\end{equation}
where the parameter $\alpha = \mathcal{O}(1)$ accounts for the fact that QuaSSis does not actually solve Eq.~\eqref{eq:Helmoltz_Ey} but the more complex set of Eqs.~\eqref{eq:df_CL_psi}--\eqref{eq:df_mat_Ez}.

\begin{table}
\centering
\begin{tabular}{|c|c|c|}
  \hline
   shape factor order & random positions & random weights \\
   \hline
    0 & $ \left(1-\frac{1}{2^d} \right) \frac{n_b^2}{N_{ppc,b}}$ & $\frac{n_b^2 \varepsilon_b^2}{3 N_{ppc,b}} $ \\
     \hline
    1 & $ \frac{4^d-3^d}{6^d} \frac{n_b^2}{N_{ppc,b}}$   & $ \frac{1}{3} \left( \frac{2 }{3} \right)^d \frac{n_b^2 \varepsilon_b^2 }{N_{ppc,b}}$ \\
      \hline
    3 &  $\frac{14564^d-10080^d}{20160^d} \frac{n_b^2}{N_{ppc,b}} $ & $ \frac{1}{3}  \left( \frac{151}{315} \right)^d \frac{n_b^2 \varepsilon_b^2 }{N_{ppc,b}}$ \\
  \hline 
\end{tabular}
    \caption{Theoretical variance of the initial beam density fluctuations $\delta \rho_b^2$ when initializing the MPs with random positions/equal weights or with regularly spaced positions/random weights.}
    \label{table:var}
\end{table}


We now assume that the transverse beam temperature is weak enough that $\delta \rho_b \gg (\Delta y/\Delta \xi)^2 \delta J_{by}^2$, and thus $\delta E_y \simeq \alpha \Delta y \sqrt{\delta \rho_b^2}$. As detailed in the Appendix, the variance $\delta \rho_b^2$ can be analytically derived depending on the MP initialization method employed (either random positions with equal weights or regular spacing with random weights). The resulting scalings, valid for both PIC and QS-PIC schemes, are given in Table~\ref{table:var} as functions of MP number, shape factor order, physical beam density and mesh dimensionality ($d$). Apart from the expected dependence on $N_{ppc,b}^{-1}$, we observe that $\delta \rho_b^2$ is proportional to $\epsilon_b^2$ when using random weighting, meaning that the initial fluctuation level can be arbitrarily tuned at a fixed computational cost.   

 \begin{figure}
    \centering
    \includegraphics[scale=0.27]{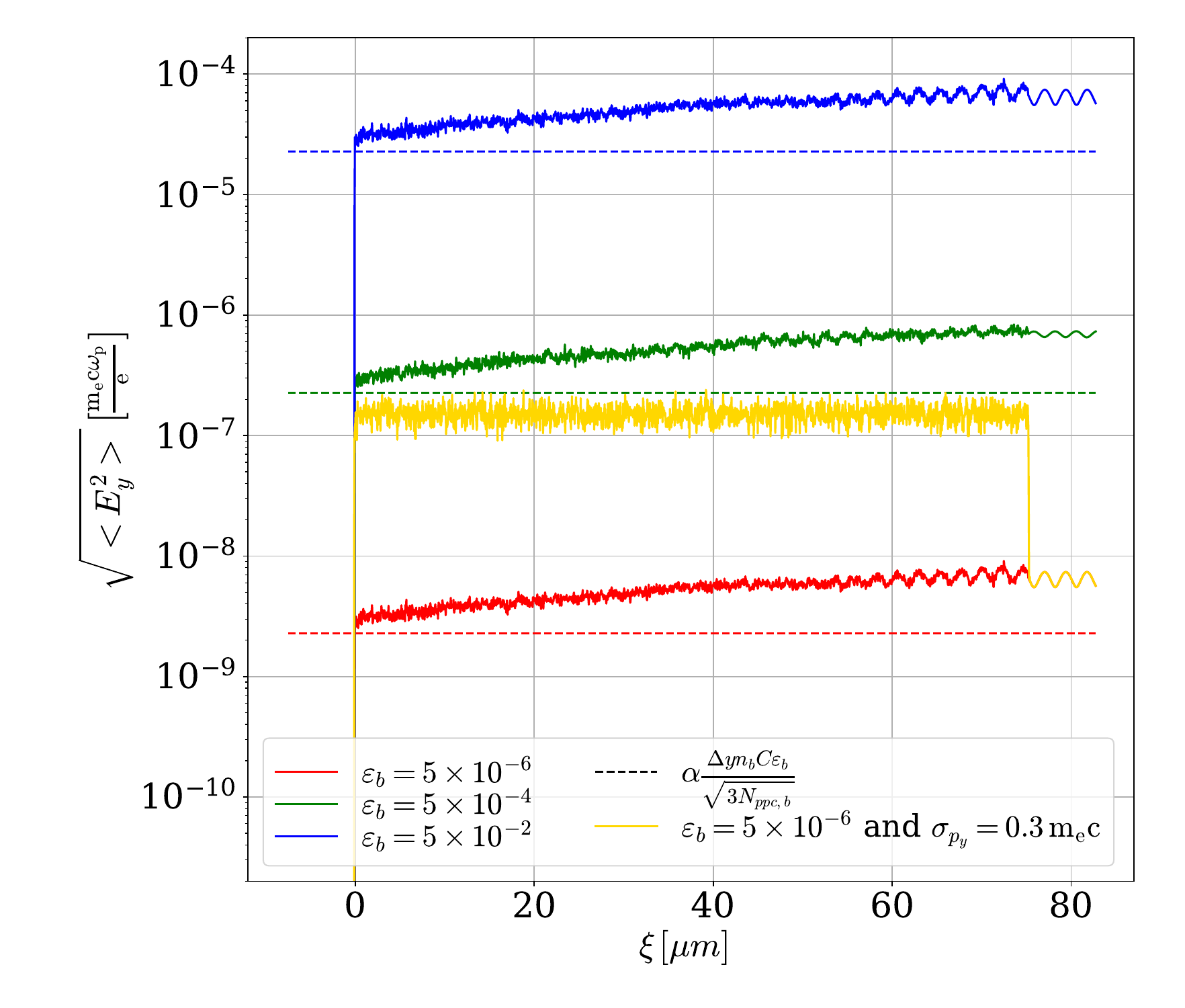}
    \caption{Initial ($s=0$) level of $E_y$ fluctuations as a function of $\xi$ for different $\varepsilon_b$ values. The quadratic mean of $E_y$ along $y$ is plotted.  Apart from $\varepsilon_b$, the simulation parameters are those used in Fig.~\ref{fig:gr_QvC}. For the yellow curve, a transverse temperature is added to the beam, characterized by a rms spread $\sigma_{p_y}/m_e c = 0.3$ in the transverse momentum distribution. The solid lines give the simulation results and the dashed line shows the fluctuation prediction given by Eq.~\eqref{eq:pred_fluc}.}
    \label{fig:fluctationEy}
\end{figure}

In particular, for our reference simulations ($d=2$, random weights, 3th order shape factors), we obtain
\begin{equation}
    \delta E_y \simeq \frac{151\alpha}{315\sqrt{3}} \frac{\Delta y n_b \varepsilon_b}{\sqrt{N_{ppc,b}}} \,,
    \label{eq:pred_fluc}
\end{equation}
To validate this formula, we repeated the simulation displayed in Fig.~\ref{fig:gr_QvC} but with varying $\varepsilon_b$ values, ranging from $5 \times 10^{-2}$ to $5 \times 10^{-6}$. Figure~\ref{fig:fluctationEy} presents the root mean square of the initial ($s=0$) electric field fluctuations along $y$ as a function of $\xi$. It is clear that the random weighting method effectively reduces the initial field fluctuations to very low levels without requiring an increase in $N_{ppc,b}$. Notably, Table~\ref{table:var} indicates that achieving a similar noise level with the random position method at $\varepsilon_b=5 \times 10^{-6}$ would necessitate an excessively high MP number ($N_{ppc,b}>2.5 \times 10^7$). Furthermore, Fig.~\ref{fig:fluctationEy} shows a good agreement between the simulated noise level and the prediction given by Eq.~\eqref{eq:pred_fluc} (plotted as dashed lines, using $\alpha=1.76$ is all cases). A slight increase in noise along $\xi$ is also observed, which we attribute to the background plasma's response to the beam, resulting in accumulated noise. However, this secondary noise remains proportional to the beam-induced fluctuations, and is therefore also controllable by $\varepsilon_b$. Additional tests (not shown) confirm the other scalings (with $n_b$, $\varepsilon_b$, $N_{ppc,b}$ and $\Delta y$) predicted by Eq.~\eqref{eq:pred_fluc}.


 
\begin{figure}
    \centering
    \includegraphics[scale=0.21]{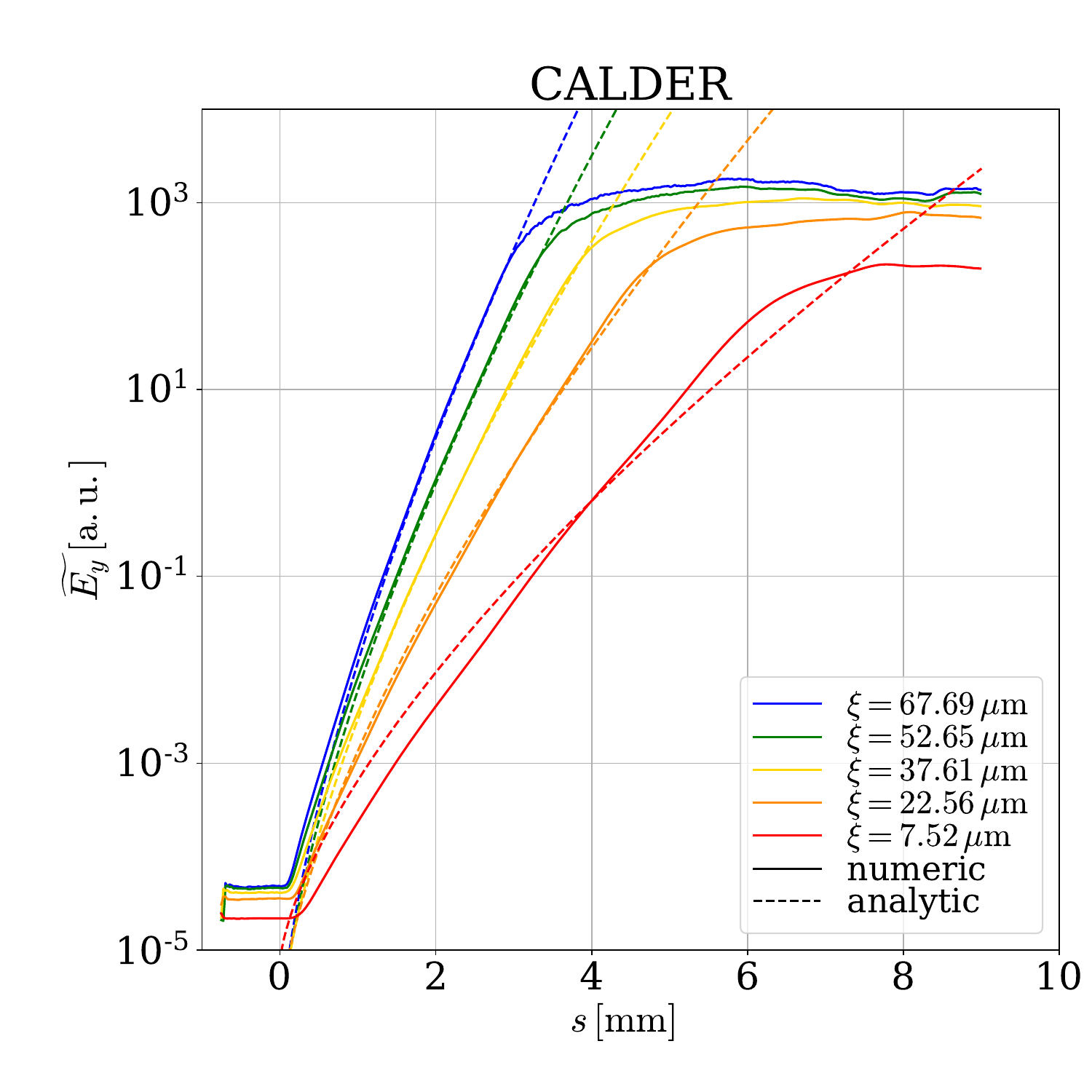}
    \includegraphics[scale=0.21]{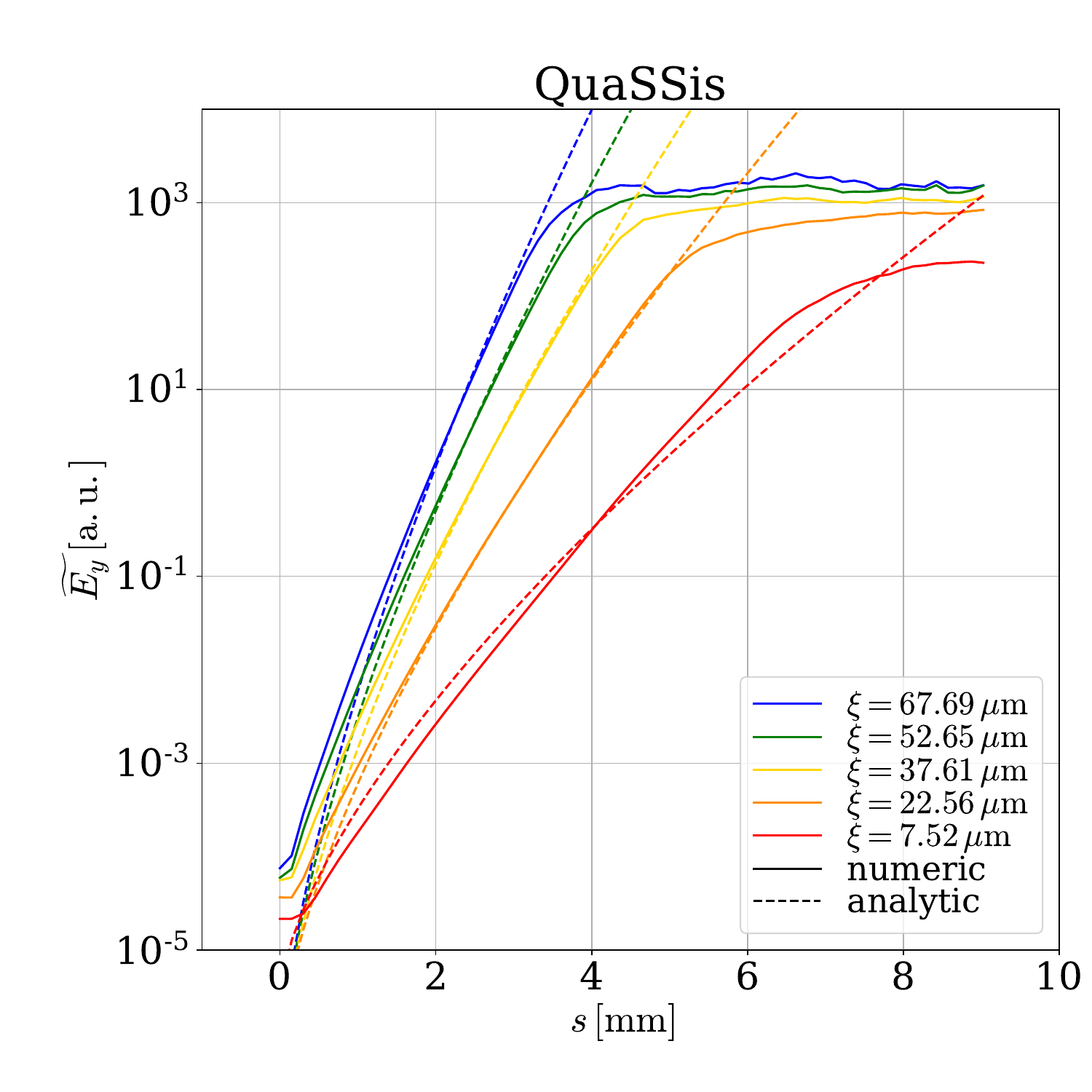}
    \caption{Spectral amplitude $\widetilde{E}_y$ as a function of the propagation distance and the position in the beam, from CALDER (left) and QuaSSis (right) simulations using the parameters of Fig.~\ref{fig:gr_QvC} except for $\varepsilon_b = 5 \times 10^{-6}$. In the CALDER simulation, the beam first propagates ballistically over $750\rm\, mm$. Dashed lines plot the theoretical fits of Eq.~\eqref{eq:Gamma_spt}. }
    \label{fig:gs}
\end{figure} 

Figure~\ref{fig:gs} shows the time histories of the spectral amplitude $\widetilde{E}_y$ as extracted from CALDER (left) and QuaSSis (right) simulations with $\varepsilon_b=5\times 10^{-6}$. A behavior similar to the results of Fig.~\ref{fig:gr_QvC} is obtained, except that the initial noise is much ($\sim 10^{-4}\times$) lower, so the beam has to propagate over a longer distance to saturate. Evidently, this also allows for an extended time span over which the linear OTSI phase is accurately captured by the theoretical prediction (dashed lines). 
In Fig.~\ref{fig:gr_QvC}, saturation happens around $s \simeq 1.3\,\rm mm$ at $\xi = 67.69\,\rm \mu m$ and $s\simeq 3.2\,\rm mm$ at $\xi = 7.52\,\rm \mu m$, while in Fig.~\ref{fig:gs}, it occurs around $s \simeq 3.1\,\rm mm$ at $\xi =  67.69\,\rm \mu m$ and $s \simeq 6.8\,\rm mm$ at $\xi = 7.52\,\rm \mu m$. However, each slice of the beam saturates at the same level in both runs.

\subsection{Competition between random weighting and temperature of beam macroparticles in noise generation}

We now consider the effect of a weak, yet finite, transverse beam temperature on the field noise. Numerically, the transverse momenta of the beam MPs are randomly drawn from a normal distribution with zero mean and standard deviation $\sigma_{p_y} \ll \gamma_b$. 
The variance $\delta J_{b y}^2$ of the beam's transverse current fluctuations can then be derived similarly to $\delta \rho_b^2$ in \ref{appendix_delta_n_b}, but it is now the transverse current of each MP ($w_i q_i p_{y,i}/\gamma_b$) that is deposited on the mesh instead of its charge $w_i q_i$. The resulting expressions are listed in Table~\ref{table:var2}.
With random weighting, $\delta J_{b y}^2$ is found to be proportional to $1 + \varepsilon_b^2/3 $. Thus, if $ \varepsilon_b \ll 1$, one has $\delta J_{b y} \propto n_b \sigma_{p_y}/(\gamma_b \sqrt{N_{ppc,b}})$, meaning that $\delta J_{by}$ is mainly determined by the MP number while the charge density noise $\delta \rho_b$ remains controlled by $\varepsilon_b$.

\begin{table}
    \centering
    \begin{tabular}{|c|c|c|}
    \hline
    shape factor order  &  random positions  & random weights \\
    \hline
    0 & $ \frac{\sigma_{p_y}^2}{\gamma_b^2}\frac{n_b^2}{N_{ppc,b}}$ & $ \left(1 + \frac{\varepsilon_b^2}{3} \right) \frac{\sigma_{p_y}^2}{\gamma_b^2} \frac{n_b^2}{N_{ppc,b}}$ \\
    \hline
    1 & $ \frac{1}{3^d} \frac{\sigma_{p_y}^2}{\gamma_b^2}\frac{n_b^2}{N_{ppc,b}}$ & $ \left( \frac{2}{3} \right)^d \left(1 + \frac{\varepsilon_b^2}{3} \right) \frac{\sigma_{p_y}^2}{\gamma_b^2} \frac{n_b^2}{N_{ppc,b}}$ \\
    \hline
    3 & $\frac{1121^d}{2520^d} \frac{\sigma_{p_y}^2}{\gamma_b^2} \frac{n_b^2}{N_{ppc,b}}$ & $ \left( \frac{151}{315} \right)^d \left(1 + \frac{\varepsilon_b^2}{3} \right) \frac{\sigma_{p_y}^2}{\gamma_b^2} \frac{n_b^2}{N_{ppc,b}}$ \\
    \hline
    \end{tabular}
    \caption{Theoretical variance of the initial beam density fluctuations $\delta J_{by}^2$ when initializing the MPs with random positions/equal weights or with regularly spaced positions/random weights.}
    \label{table:var2}
\end{table}

For our reference simulations ($d=2$, random weighting, 3rd order shape factors), Eq.~\eqref{eq:Approx_Ey} then becomes
\begin{equation}
    \delta E_y = \frac{151 \alpha}{315} \frac{\Delta y n_b}{\sqrt{N_{ppc,b}}} \left( \frac{\varepsilon_b^2}{3}  +  \frac{\Delta y^2}{\Delta \xi^2}\frac{\sigma_{p_y}^2}{\gamma_b^2} \right)^{1/2} \,.
\end{equation}
This formula shows that the $E_y$ noise will be controlled by $\varepsilon_b$ as long as $\varepsilon_b > \sqrt{3}(\Delta y/\Delta \xi)\sigma_{p_y}/\gamma_b$. Otherwise, $\delta E_y$ will mainly depend on the physical momentum spread $\sigma_{p_y}$. This condition sets a limit to the random weighting method because the noise cannot be reduced further. The yellow curve in Fig.~\ref{fig:fluctationEy} illustrates this limiting effect: with $\gamma_b = 20\,000$, $\sigma_{p_y}/m_e c=0.3$, $\Delta \xi = \Delta y$ and $\varepsilon_b=5\times10^{-6}$, we have 
$(\Delta y/\Delta \xi) \sigma_{p_y}/\gamma_b = 1.5\times10^{-5} > \varepsilon_b$. Thus, the transverse beam temperature accounts for the observed $E_y$ fluctuation level, which is $\sim 50\times$ higher than in the case of $\sigma_{p_y}=0$ and $\varepsilon_b=5\times10^{-6}$ (red curve).

\section{Conclusions}

We have demonstrated that the quasistatic PIC method can efficiently simulate the propagation of ultrarelativistic dilute particle beams subject to the oblique two-stream instability. To achieve this, we have developed a new 2D QS-PIC code, QuaSSis, which incorporates several established schemes along with two key improvements.
First, it features an original treatment of transverse boundary conditions, allowing one to model effectively infinite transverse density profiles (of academic or astrophysical interest) and benchmark the simulation results against theoretical models.
Second, we have implemented a macroparticle initialization method based on random weights to tune, and possibly reduce by orders of magnitude, the level of electromagnetic fluctuations in the beam-plasma system. With this method, realistically low noise levels, such as those expected in accelerator experiments, can be reproduced so that the simulation can provide reliable predictions for the interaction length before the instability saturates.

Thanks to these developments, very good agreement has been obtained between QuaSSis simulations and theoretical predictions of the perturbative spatiotemporal phase of the OTSI \cite{PSMC_2022}. QuaSSis also closely reproduces the results of standard PIC simulations, with both infinite and Gaussian transverse beam profiles. Above all, the QuaSSis results confirm a decisive advantage of the QS-PIC method already exploited for LWFA and PWFA studies: the simulation time can be reduced by up to three orders of magnitude compared to full PIC simulations, which can significantly facilitate future studies of ultrarelativistic beam-plasma instabilities.


\section*{Acknowledgments}

We acknowledge the Grand Equipement National de Calcul Intensif GENCI-TGCC for providing us access to the supercomputer IRENE under Grant No. A0170512993. This work has been supported by the Agence Nationale de la Recherche (ANR) (UnRIP project, Grant No. ANR-20-CE30-0030).

\appendix

\section{Initial density fluctuations with random weighting}
\label{appendix_delta_n_b}

Here we outline the method used to derive the formulas listed in Table~\ref{table:var} for the variance of the MP density fluctuations. 

In a PIC simulation, the charge density of a given plasma species is computed at mesh node $k$ as
\begin{equation} 
    n_k = \frac{1}{V} \sum_{i=1}^{N_p} w_i \mathcal{S} (\pmb{x}_i, \pmb{x}_k ) \,.
\end{equation} 
We have introduced $N_p$ the total number of MPs, $\pmb{x}_i$ and $w_i$ the position and weight of MP $i$, $\mathcal{S}$ the shape factor, $\pmb{x}_k$ the position of the node and $V$ the cell volume. 

As an example, we consider the case of a zero-order shape factor in 1D ($d=1$):
\begin{equation} 
    \mathcal{S}^0 (x_i, x_k, \Delta x) \longrightarrow
    \begin{cases}
        1, & \text{if } \frac{\vert x_k - x_i \vert}{\Delta x} \leq \frac{1}{2} \,, \\
        0, & \text{otherwise}.
    \end{cases} 
\end{equation}
where $\Delta x$ is the cell size and $x_k=k \Delta x$. We now address the two MP initialization methods considered in the main text. 

\paragraph{Random positioning with equal weights}

Let us consider that MP $i$ is located in cell $k$. Its weight and position are initialized as
\begin{align} 
    &w_i = \frac{V n}{N_{ppc}} \,, 
    \label{eq:w_rp} \\
    &x_i \hookrightarrow \mathcal{U}\left[x_k, x_k + \Delta x \right] \,,
    \label{eq:x_rp}
\end{align} 
where $n$ is the prescribed density, and $N_{ppc}$ is the MP number per cell. Here, the MP position, $x_i$, is a random variable with its probability distribution given by $\mathcal{U}$, a continuous uniform distribution equal to 1 in the interval $\left[x_k, x_k + \Delta x \right]$ and 0 outside of it.

Before calculating the variance of $n_k$, we assume that the prescribed beam density and MP number are the same in all cells. In 1D, the zero-order shape factor at node $k$ is positive only over the segment $[ x_k - \Delta x/2, x_k + \Delta x/2 ]$, so only the MPs from the two cells adjacent to node $k$ should be considered. If a MP is initialized in cell $k$ (i.e., $x_k < x_i < x_{k+1}$), the probability that its position $x_i$ lies within $[ x_k - \Delta x/2, x_k + \Delta x/2 ]$ is binary: either $x_i \in \left[x_k, x_k + \Delta x/2\right]$ and $\mathcal{S}^0 (x_i, x_k, \Delta x) = 1$, or $x_i \in \left[x_k + \Delta x/2, x_{k+1}\right]$ and $\mathcal{S}^0 (x_i, x_k, \Delta x) = 0$. Since the random $x_i$ obeys Eq.~\eqref{eq:x_rp}, the shape factor result follows a Bernoulli distribution: 
\begin{equation}
    \mathcal{S}^0 (x_i, x_k, \Delta x) \hookrightarrow \mathcal{B}(1/2) \,. 
    \label{eq:Bernoulli_1D}
\end{equation}
Of course, the same result is obtained if MP $i$ lies in cell $k-1$ (i.e., $x_{k-1} < x_i < x_k$) due to the symmetry of the shape factor.

Thus, summing over all MPs, the density at cell $k$ obeys the random law
\begin{equation} 
    n_k \hookrightarrow \frac{w_i}{V} \left[ \sum_{i=(k-1)N_{ppc}+1}^{kN_{ppc}} \mathcal{B}(1/2) + \sum_{i=kN_{ppc}+1}^{(k+1) N_{ppc}} \mathcal{B}(1/2) \right] \,. 
\end{equation}
Using Eq.~\eqref{eq:w_rp}, we can rewrite $n_k$ as a binomial distribution: 
\begin{equation} 
    n_k \hookrightarrow \frac{n}{N_{ppc}} \mathcal{B}(2N_{ppc}, 1/2) \,. 
\end{equation}
Its average value is $\mathbb{E}[n_k] = n$ as expected, and its variance is 
\begin{equation} 
    \mathbb{V}[n_k] = \frac{n^2}{2 N_{ppc}} \,. 
\end{equation}

Generalizing this formula to a multidimensional geometry is straightforward. Introducing the position vector $\pmb{x}$ with components $\{x_j \}_{1 \le j \le d}$, the multidimensional shape factor is defined by
\begin{equation} 
    \mathcal{S}(\pmb{x}_i, \pmb{x}_k, \pmb{\Delta x}) = \prod_{j=1}^d \mathcal{S} (x_{i,j} \,, x_{k,j}, \Delta x_j) \,,
    \label{eq:shape_factor_dd}
\end{equation} 
where $\pmb{\Delta x}$ is the vector representing the spatial step in each dimension, $k_j$ and $x_{k,j}$ are respectively the node number and position in the $j$th direction.
For the zero-order shape factor, the use of Eq.~\eqref{eq:Bernoulli_1D} leads to
\begin{equation} 
    \mathcal{S}^0(\pmb{x_i}, \pmb{x}_k, \pmb{\Delta x}) \hookrightarrow \prod_{j=1}^d \mathcal{B}(1/2) \,.
\end{equation}
With MPs coming from the $2^d$ cells surrounding node $\pmb{k}$, we finally get
\begin{equation} 
    n_{\pmb{k}} \hookrightarrow \frac{n}{N_{ppc}} \mathcal{B}(2^d N_{ppc}, 1/2^d) \,, 
\end{equation}
leading to
\begin{align} 
    &\mathbb{E}[n_{\pmb{k}}] = n \,, \\
    &\mathbb{V}[n_{\pmb{k}}] = \frac{n^2}{N_{ppc}} \left( 1 - \frac{1}{2^d} \right) \,. 
\end{align}
The latter formula is reported in Table~\ref{table:var}, where the prescribed density and MP number are taken to be those of the beam.

The same method can be applied to higher-order shape factors, but the details of the calculation are not presented here for the sake of simplicity. The results for the first- and third-order shape factors are given in Table~\ref{table:var} for the beam parameters.

\paragraph{Regular spacing with random weights}

The MP positions are now set to
\begin{equation} 
    x_i = x_k + \left(i - \frac{1}{2}\right)\frac{\Delta x}{N_{ppc}} \,, 
    \label{eq:x_rw} 
\end{equation}
for $kN_{ppc} \le i \le (k+1) N_{ppc}$. In this case, should the MPs be assigned the same weight, one would have exactly $n_k = n$ with no fluctuation on the mesh. Instead, we take the MP weights to be
\begin{equation} 
    w_i \hookrightarrow \frac{V n}{N_{ppc}} \left( 1 + \varepsilon \mathcal{U} \left[-1,1\right] \right) \,. 
    \label{eq:w_rw} 
\end{equation} 
Due to the known MP positions within the cell, we have $\mathcal{S}^0 (x_i, x_k, \Delta x) = 1$ for the first $N_{ppc}/2$ MPs that verify $x_k < x_i < x_{k+1}$, and $\mathcal{S}^0 (x_i, x_k, dx) = 0$ for the remaining MPs (assuming an even number of MPs). The same applies to MPs between $x_{k-1}$ and $x_k$. Therefore, $n_k$ is given by
\begin{equation} 
    n_k = \frac{1}{V} \left[
    \sum_{i=\left(k-\frac{1}{2}\right) N_{ppc}+1}^{k N_{ppc}} w_i
    + \sum_{i = k N_{ppc} + 1 }^{\left(k+\frac{1}{2}\right) N_{ppc}} w_i  \right] \,.
\end{equation}
Using the random weight from Eq.~\eqref{eq:w_rw}, we have 
\begin{equation} 
    n_k \hookrightarrow \frac{n}{N_{ppc}} \sum_{i=1}^{N_{ppc}} \left(1 + \varepsilon \mathcal{U}\left[-1,1\right]\right) \,.
    \label{eq:nk_RW}
\end{equation}
The mean and variance of $n_k$ are given by
\begin{align} 
    &\mathbb{E}[n_k] = n \,, \\ 
    &\mathbb{V}[n_k] = \frac{n^2 \varepsilon^2}{3 N_{ppc}} \,.
    \label{eq:Vnk_RW}
\end{align}

In $d$ dimensions, only $N_{ppc}/2^d$ MPs in each cell surrounding node $\pmb{k}$ will contribute to the charge deposited on this node. However, this is compensated by the fact that $2^d$ cells surround the node, so Eqs.~\eqref{eq:nk_RW}--\eqref{eq:Vnk_RW} remain valid regardless of the dimensionality. This result is reported in Table~\ref{table:var} for the beam parameters, along with its generalization to other shape factor orders.






\bibliographystyle{elsarticle-num}
\bibliography{ref}







\end{document}